\documentclass[11pt,a4paper]{article}
\pdfoutput=1
\usepackage[pdftex]{graphicx}
\usepackage{amsmath}
\usepackage{amssymb}
\usepackage{graphicx}
\usepackage{theorem}
\usepackage{array}
\usepackage[numbers, sort]{natbib}
\usepackage{hyperref}
\usepackage{color}
\usepackage{thumbpdf}
\usepackage[toc, page]{appendix}
\usepackage[footnotesize]{caption}

\numberwithin{equation}{section}

\newcommand{\be}{\begin{equation}}
\newcommand{\ee}{\end{equation}}
\newcommand{\beq}{\begin{equation}}
\newcommand{\eeq}{\end{equation}}
\newcommand{\bea}{\begin{eqnarray}}
\newcommand{\eea}{\end{eqnarray}}
\newcommand{\ox}{\omega}

\def\imag{\,{\rm Im}\, }
\def\real{\,{\rm Re}\, }

\allowdisplaybreaks[2]

\makeatletter

\newcommand{\Rmnum}[1]{\expandafter\@slowromancap\romannumeral #1@}
\makeatother

\begin{document}

\title{\huge \bf{Neutrino Mixing Angles in Sequential Dominance to NLO and NNLO}}
\author{
    S.~Antusch,$^{1,}$\footnote{E-mail: antusch@mppmu.mpg.de}~
    S.~Boudjemaa$^{2,}$\footnote{E-mail: sb26@soton.ac.uk} ~and
    S.~F.~King$^{2,}$\footnote{E-mail: sfk@hep.phys.soton.ac.uk} \vspace{2mm}\\
    $^1$ Max-Planck-Institut f\"ur Physik (Werner-Heisenberg-Institut),\\
    F\"ohringer Ring 6, D-80805 M\"unchen, Germany\\
    $^2$ School of Physics and Astronomy, University of Southampton,\\
    SO17 1BJ Southampton, United Kingdom}
\date{}
\maketitle

\begin{abstract}
\noindent
Neutrinos with hierarchical masses and two large mixing angles may naturally originate from sequential dominance (SD).
Within this framework we present analytic expressions
for the neutrino mixing angles including
the next-to-leading order (NLO) and
next-to-next-to-leading order (NNLO) corrections arising from the second lightest and lightest neutrino
masses. The analytic results for neutrino mixing angles in SD
presented here, including the NLO and NNLO corrections, are applicable to
a wide class of models and may provide useful insights when
confronting the models with data from high precision neutrino experiments.
We also point out that for special cases of SD corresponding to form dominance (FD)
the NLO and NNLO corrections both vanish. For example
we study tri-bimaximal (TB) mixing via constrained sequential dominance (CSD)
which involves only a NNLO correction and tri-bimaximal-reactor
(TBR) mixing via partially constrained sequential dominance (PCSD)
which involves a NLO correction suppressed by the small reactor angle and show that the
analytic results
have good agreement with the numerical results for these cases.

\end{abstract}

\newpage

\section{Introduction}

The flavour problem of the Standard Model (SM), i.e.\ the origin of the observed fermion masses and mixings, is one of the greatest unsolved puzzles in particle physics. The discovery of neutrino masses and large lepton mixing angles has added another piece to this puzzle. A successful theory of flavour has to explain why neutrino masses are small and why the mixing in the lepton sector is large whereas the mixing in the quark sector is small.

A promising approach to both issues is provided by the type \Rmnum{1} see-saw mechanism \cite{seesaw} with right-handed neutrino dominance \cite{King:1998jw}. 
In particular in this paper we shall focus on the sequential dominance (SD)
of three right-handed neutrinos \cite{King:1999mb,Lavignac:2002gf,King:2002nf,King:2002qh,King:2004,Haba:2008dp}.
While the seesaw mechanism is generically realised in left-right symmetric Grand Unified Theories (GUTs) and can explain the smallness of neutrino masses, SD, i.e.\ the concept that the right-handed neutrinos contribute to the neutrino mass matrix via the see-saw mechanism with sequential strength, provides insight into how large neutrino mixing can arise in a natural way for a hierarchical neutrino mass spectrum $|m_3|>|m_2|\gg |m_1|$ where experimentally $|m_3|\sim 5|m_2|\sim 0.05$ eV \cite{Bandyopadhyay:2007kx}. According to SD, the leading order (LO) contribution to the neutrino mass matrix comes from one single right- handed neutrino resulting in a single neutrino mass, namely the heaviest neutrino mass $m_3$, and the ``atmospheric'' mixing angle $\theta_{23}$ \cite{King:1998jw}. The second largest next-to-leading order (NLO) contribution to the neutrino mass matrix in SD, arising from a second right-handed neutrino, induces the second heaviest neutrino mass $m_2$ as well as the ``solar'' and ``reactor'' mixing angles $\theta_{12}$ and $\theta_{13}$, respectively \cite{King:1999mb}. 
The third largest next-to-next-to-leading order (NNLO) contribution to the neutrino mass matrix in SD, arising from a third right-handed neutrino is responsible for the lightest neutrino mass $m_1$ and this is usually considered to be sufficiently small so that the NNLO corrections do not perturb too much the determination of neutrino mixing angles in SD
\cite{King:1999mb,Lavignac:2002gf,King:2002nf,King:2002qh,King:2004}.

The analytic estimates of the mixing
angles in SD have so far only been presented to LO 
\cite{King:1999mb,Lavignac:2002gf,King:2002nf,King:2002qh,King:2004}.
However the NLO corrections
of order $m_2/m_3$ might be expected to introduce mass dependent corrections of order
20\%
to the analytic expressions for the neutrino mixing angles.
Given the present and planned program of neutrino experiments, it is apparent that
neutrino
physics is now entering the precision era, with all mixing angles being measured in the
forseeable
future to an accuracy of a few per cent \cite{Bandyopadhyay:2007kx}. Against such
experimental progress it is clear that
theoretical errors of order 20\% are no longer acceptable.
This motivates a study of NLO corrections to SD.

Typically unified models predict a third
right-handed neutrino which contributes to the seesaw mechanism at NNLO and
provides a mass $m_1$ to the lightest neutrino which will also give some corrections
to the analytic expressions for the neutrino mixing angles at NNLO.
The importance of NNLO corrections to the neutrino mixing angles of order $m_1/m_3$
clearly depends on the size of $m_1$.
However, in unified models with SD, the value of $m_1$ is governed
by a rather large third family Yukawa coupling
to the third right-handed neutrino of order unity, and in such models the NNLO corrections
could be significant. This motivates studying NNLO corrections to SD in addition to the NLO corrections.

We shall refer to the NNLO corrections as ``soft'' since they vanish
in the strongly hierarchical vanishing $m_1$ limit.
By contrast we call the NLO corrections ``hard'' since it is not possible
to decouple these corrections by taking the vanishing limit of $m_2$.
However there is a class of SD theories in which the NLO and NNLO corrections both
vanish namely those which satisfy form dominance (FD) \cite{Chen:2009um}
(see also \cite{Antusch:2007jd}). FD \cite{Chen:2009um} is a criterion whereby the
columns of the Dirac matrix
are proportional to the respective columns of the
neutrino mixing matrix in a basis
where the charged lepton and heavy Majorana mass matrix are diagonal.
FD implies that the neutrino mass matrix resulting from the type I see-saw
mechanism is form diagonalizable with the
neutrino mixing angles being independent of the neutrino masses.
This observation helps to explain why the LO results for SD are often
observed to work better than expected in particular cases.
For example there are special cases where FD is only
violated by soft corrections leading to NNLO corrections but not NLO corrections,
namely those which satisfy constrained sequential dominance (CSD) \cite{King:2005bj}
where a strong hierarchy is assumed and terms proportional to $m_1$ violate FD.

As the precision of the neutrino data has improved, it has
become apparent that lepton mixing is consistent with the so called tri-bimaximal (TB)
mixing
pattern \cite{tribi}
\begin{eqnarray}
U_{TB} =
\left( \begin{array}{rrr}
\sqrt{\frac{2}{3}}  & \frac{1}{\sqrt{3}} & 0 \\
-\frac{1}{\sqrt{6}}  & \frac{1}{\sqrt{3}} & \frac{1}{\sqrt{2}} \\
\frac{1}{\sqrt{6}}  & -\frac{1}{\sqrt{3}} & \frac{1}{\sqrt{2}}
\end{array}
\right).
\label{MNS0}
\end{eqnarray}
TB lepton mixing in particular hints at a spontaneously broken family symmetry $G_f$
which might underpin a flavour theory of all quarks and leptons, but which might only
reveal itself in the neutrino sector.
What is the nature of such a family symmetry?
In the (diagonal) charged lepton mass basis,
it has been shown that the neutrino mass matrix leading to TB mixing is invariant
under the Klein group\cite{Lam}.
The observed neutrino flavour symmetry of the Klein group
may arise either {\it directly} or {\it indirectly} from certain classes of discrete family symmetries \cite{King:2009ap}.
Several models have been constructed that account for the
structure of leptonic mixings, e.g. \cite{S3-L,Dn-L,A4-L,S4-L,delta54-L}, while other models
extend the underlying family symmetry to provide
a description of the complete fermionic structure
\cite{S3-LQ,Dn-LQ,Q6-LQ,A4-LQ,A4-LQ:Morisi,A4-LQ:King,doubleA4-LQ,S4-LQ:Hagedorn,S4-group,S4-LQ,Mohapatra:2003tw,King:2009mk,A4-SU5,Z7Z3-LQ,T7-LQ,delta27-LQ:King,SO(3)-LQ:King,SU(3)-LQ:Ross}.
Most of these models satisfy the conditions of either FD \cite{Chen:2009um}
or CSD \cite{King:2005bj}. In particular those in
\cite{A4-LQ,A4-LQ:Morisi,doubleA4-LQ,S4-LQ:Hagedorn,S4-group,S4-LQ,A4-SU5}
satisfy FD while those in
\cite{A4-LQ:King,Mohapatra:2003tw,King:2009mk,Z7Z3-LQ,delta27-LQ:King,SO(3)-LQ:King,SU(3)-LQ:Ross}
satisfy CSD. We emphasise that FD and CSD are not only compatible with symmetry but are actually 
enforced by symmetry in these models. For example the texture zeros elements of the Dirac
mass matrix that we shall encounter in Section 6 are all enforced by symmetries in realistic models.

On the other hand there are many other models in the literature too numerous to mention
that do not yield very precise TB mixing as a result of a symmetry for example,
see \cite{Reviews} for review papers with more extensive references.
Many of these other models will satisfy the looser conditions of SD
\cite{King:1998jw,King:1999mb,King:2002nf,King:2002qh,King:2004} but not those of CSD or FD.
Such general SD models would become necessary if significant deviations from TB mixing are observed.
For example, while TB mixing implies zero mixing angle $\theta_{13}$, the most recent experimental
results hint at
a non-zero (and in fact comparatively large) reactor mixing in the one sigma range
$s_{13}^2=\sin^2\theta_{13} \approx 0.02 \pm 0.01$ \cite{Fogli:2009ce}
corresponding to a reactor angle of $\theta_{13}\approx 8^\circ \pm
2^\circ$.
Assuming TB mixing of the neutrino mass matrix via CSD, various effects can generate
deviations
from $\theta_{13}=0$, for example charged lepton corrections \cite{sumrule}, RG running
effects
\cite{RG,Dighe:2006sr,Boudjemaa:2008jf} or corrections from canonical normalisation
\cite{Antusch:2007ib,cnorm}.
However, by appealing to these effects, it is difficult to understand why the reactor angle
should be
much larger than the deviation of the solar angle from its tri-bimaximal value.
The simplest possibility for allowing a sizeable non-zero reactor angle (e.g. $
\theta_{13}\approx
8^\circ$) while maintaining accurately the TB predictions
for the solar and atmospheric angles, $\theta_{12}\approx 35^\circ$ and $
\theta_{23}\approx 45^\circ$,
is called tri-bimaximal-reactor (TBR) mixing \cite{King:2009qh} corresponding to the mixing
matrix,
\begin{eqnarray}
U_{TBR} =
\left( \begin{array}{ccc}
\sqrt{\frac{2}{3}}  & \frac{1}{\sqrt{3}} & \frac{1}{\sqrt{2}}re^{-i\delta } \\
-\frac{1}{\sqrt{6}}(1+ re^{i\delta })  & \frac{1}{\sqrt{3}}(1- \frac{1}{2}re^{i\delta })
& \frac{1}{\sqrt{2}} \\
\frac{1}{\sqrt{6}}(1- re^{i\delta })  & -\frac{1}{\sqrt{3}}(1+ \frac{1}{2}re^{i\delta })
 & \frac{1}{\sqrt{2}}
\end{array}
\right),
\label{MNS3}
\end{eqnarray}
where we have introduced the reactor parameter $r$ defined by $s_{13} = \frac{r}{\sqrt{2}}$
\cite{King:2007pr}
where $s_{13}^2 \approx 0.02$ corresponds to $r\approx 0.2$.
TBR mixing can arise
from type \Rmnum{1} see-saw mechanism via a very simple modification to CSD called
partially constrained sequential dominance (PCSD) \cite{King:2009qh}. Estimates suggest
that PCSD is capable of
accommodating a
sizeable reactor angle while the atmospheric and solar angles are predicted to remain
close to
their TB values \cite{King:2009qh}.
However, as in the case of CSD, the analytic results for the mixing angles have only been
presented to LO, and once again the NLO corrections of order $m_2/m_3$ might be
expected to introduce mass
dependent corrections of order 20\%. However we shall show that for the case of PCSD
such NLO corrections
become suppressed by $r$ leading to small corrections, for example 4\% correction for $r
\approx 0.2$.
We shall drop such corrections in our analytic results, though they are apparent in the
numerical results.

In this paper, then, within the general framework of the type I see-saw mechanism with SD, we
derive approximate analytic
formulae of the neutrino mixing angles including both the hard NLO and the soft NNLO
corrections. These results may be applied in a model independent way to a very large class of
theories which satisfy SD
away from the FD limit. The derivation of these
analytic expressions builds on the results presented in \cite{King:2002nf}
where the NLO and NNLO corrections were not considered \footnote{Although the NLO
corrections
were calculated for the atmospheric angle they were not considered for the other angles,
and NNLO
corrections were completely neglected \cite{King:2002nf}.}. We apply these analytic results
to
the cases of CSD and PCSD, as examples where soft and hard violations of FD are
expected, respectively.
However in PCSD the hard NLO corrections are suppressed by the small reactor angle and
we drop such corrections
in our analytic formulae.
We compare the analytic results to an exact numerical diagonalization of the neutrino mass
matrix, for the examples of two numerical GUT models previously studied \cite{Boudjemaa:2008jf}
with CSD and PCSD, and obtain good agreement which support our conclusions based on
the analytic formulae.

The paper is organised as follows. In Section \ref{D}, we discuss lepton mixing from
the special types of SD corresponding to CSD and PCSD as examples of soft and hard
violations
of FD leading to NNLO and NLO corrections respectively.
In Section \ref{SD}, we briefly review general SD and the LO analytic expressions for
neutrino mixing angles from \cite{King:2002nf} away from the FD limit.
Section \ref{cor} extends these results to
include NLO and NNLO corrections in terms of the NLO parameters $\epsilon_i$ and the
NNLO parameters
$\eta_i$. In Section \ref{pcsd}
we apply the analytic results to the cases of CSD and PCSD and show that the corrections
correspond to soft NNLO corrections and hard NLO corrections suppressed by the reactor
angle, respectively.
In Section \ref{numerical}, we present numerical results for the mixing angles and the
neutrino
masses using the Mathematica package
MPT/REAP
\footnote{Mixing Parameter Tools (MPT) is a package provided with REAP and it
is mainly used to extract neutrino mixing parameters.}
\cite{Antusch:2005gp} for some examples of CSD and PCSD which support our
conclusions based on the
analytic formulae.
Section \ref{conc} concludes the paper. The detailed procedure for the diagonalisation of
the left-handed
neutrino mass matrix as well as details of the derivations of the analytic results are given in
the
Appendices.

Given the large number of acronyms introduced, it may be convenient to summarise them:

SM = Standard Model

GUT = Grand Unified Theory

LO = Leading Order

NLO = Next-to Leading Order

NNLO = Next-to-Next-to Leading Order

TB = Tri-Bimaximal

TBR = Tri-Bimaximal-Reactor

SD = Sequential Dominance

LSD = Light Sequential Dominance

HSD = Heavy Sequential Dominance

FD = Form Dominance

CSD = Constrained Sequential Dominance

PCSD = Partially Constrained Sequential Dominance

REAP = Renormalisation group Evolution of Angles and Phases

MPT = Mixing Parameter Tools

\section{Lepton mixing in special cases of sequential dominance}\label{D}
The mixing matrix in the lepton sector, the MNS matrix
$U_{\mathrm{MNS}}$, is defined as the matrix which appears in the
electroweak coupling to the $W$ bosons expressed in terms of lepton
mass eigenstates. With the masses of charged leptons $M_\mathrm{e}$
and neutrinos $m_{\nu}$ written as
\begin{eqnarray}
{\cal L}=-  \bar{e}_L M_\mathrm{e} e_R
- \tfrac{1}{2}\bar{\nu}_L m_{LL} \nu_\mathrm{L}^c
+ \text{H.c.}\; ,
\end{eqnarray}
and performing the transformation from flavour to mass basis by
 \begin{eqnarray}\label{eq:DiagMe}
V_{\mathrm{e}_\mathrm{L}} \, M_{\mathrm{e}} \,
V^\dagger_{\mathrm{e}_\mathrm{R}} =
\mbox{diag}(m_e,m_\mu,m_\tau)
 , \quad
V_{\nu_\mathrm{L}} \,m_{LL}\,V^T_{\nu_\mathrm{L}} =
\mbox{diag}(m_1,m_2,m_3),
\end{eqnarray}
the MNS matrix is given by
\begin{eqnarray}\label{Eq:MNS_Definition}
U_{\mathrm{MNS}} = V_{e_\mathrm{L}} V^\dagger_{\nu_\mathrm{L}}\; .
\end{eqnarray}

In this paper we shall choose a flavour basis in which the charged lepton mass matrix is
diagonal. In this basis the MNS matrix arises from the neutrino sector, and the effective
neutrino
mass matrix $m_{LL}$ is given in terms of the neutrino masses
$m_1,m_2,m_3$ by,
\bea
m_{LL} & = &
U_{\mbox{\scriptsize MNS}}{\rm diag}
(m_{1}, \; m_{2}, \; m_{3})U_{\mbox{\scriptsize MNS}}^{T} \nonumber \\
& = &
m_{1} \Phi_{1}\Phi_{1}^{T} + m_{2} \Phi_{2}\Phi_{2}^{T} + m_{3} \Phi_{3}\Phi_{3}^{T} ,
\label{expansion}
\eea
where we have written the mixing matrix in terms of three column vectors
\beq
U_{\mbox{\scriptsize MNS}}=(\Phi_1,\Phi_2,\Phi_3).
\label{columns}
\eeq

Turning to the type I see-saw mechanism, the starting point is a heavy right-handed
Majorana
neutrino mass matrix
$M_{RR}$ and a Dirac neutrino mass matrix (in the left-right convention) $M_{LR}$, with
the light effective left-handed Majorana
neutrino mass matrix $m_{LL}$ given by the type I see-saw formula
(up to an irrelevant minus sign which henceforth is dropped),
\begin{equation}\label{eq:meff}
m_{LL} = M_{LR} M_{RR}^{-1} M_{LR}^{T}.
\end{equation}
In a basis in which $M_{RR}$ is diagonal, we may write,
\begin{equation}
M_{RR} = \mbox{diag}(Y, X, X')
\end{equation}
and $M_{LR}$ may be written in terms of three general column vectors $A,B,C$,
\begin{equation}
M_{LR} = (A,B,C).
\end{equation}
The see-saw formula then gives,
\begin{equation}
\label{eq:seesawmeff}
m_{LL} =
 \frac{AA^{T}}{Y} + \frac{BB^{T}}{X} +\frac{CC^{T}}{X'}.
\end{equation}

By comparing Eq.~(\ref{eq:seesawmeff}) to (\ref{expansion}) it is clear that
the neutrino mass matrix is form diagonalizable if we assume that the columns of the
mixing matrix $\Phi_1, \Phi_2, \Phi_3$ are proportional to the columns of the Dirac mass
matrix
$A=e\Phi_3 ,B=b\Phi_2 ,C=c'\Phi_1$, where $e,b,c'$ are real parameters,
an assumption known as form dominance (FD) \cite{Chen:2009um}
(see also \cite{Antusch:2007jd}). In this case the physical neutrino masses are given
by $m_3=e^2/Y$, $m_2=b^2/X$, $m_1={c'}^2/X'$ and the neutrino mass matrix is
diagonalized
precisely by $U_{MNS}$ due to the unitarity relation
$\Phi_i^{\dagger}\Phi_j= \delta_{ij}$.
There are no NLO or NNLO corrections in FD since the mixing matrix is determined by
the column vectors $\Phi_i$ which are independent of the parameters which determine the
neutrino masses $m_i$.
This conclusion is independent of the choice of mixing matrix
$U_{MNS}$, however the usual assumption is that it is of the TB form.

For example, TB mixing in Eq.~(\ref{MNS0}) results from
\begin{equation}
\label{FD2}
A  = \frac{\tilde e}{\sqrt{2}}
\left(
\begin{array}{r}
0 \\
1 \\
1
\end{array}
\right),
\ \
B = \frac{\tilde b}{\sqrt{3}}
\left(
\begin{array}{r}
1 \\
1 \\
-1
\end{array}
\right),\ \
C = \frac{\tilde c'}{\sqrt{6}}
\left(
\begin{array}{r}
2 \\
-1 \\
1
\end{array}
\right),
\end{equation}
where we have written $A= \tilde e \Phi^{TB}_3$, $B= \tilde b \Phi^{TB}_2$, $C= \tilde c'
\Phi^{TB}_1$,
with $U_{\mbox{\scriptsize TB}}=(\Phi^{TB}_1,\Phi^{TB}_2,\Phi^{TB}_3)$
and we identify the physical neutrino
mass eigenvalues as $m_3=\tilde e^2/Y$, $m_2=\tilde b^2/X$, $m_1=\tilde c'^2/X'$.

It is interesting to compare FD to
Constrained Sequential Dominance (CSD) defined in
\cite{King:2005bj}. In CSD a strong hierarchy $|m_1|\ll |m_2| < |m_3|$ is assumed
which enables $m_1$ to be effectively ignored (typically this is achieved by
taking the third right-handed neutrino mass $X'$ to be very heavy leading to a very light
$m_1$).
Thus CSD is seen to be just a special case of FD corresponding to a strong neutrino mass
hierarchy. FD on the other hand is more general
and allows any choice of neutrino masses including
a mild hierarchy, an inverted hierarchy or a quasi-degenerate mass pattern.

In practice, CSD is defined by
only assuming the first and second conditions in Eq.~(\ref{FD2}) \cite{King:2005bj},
with the third condition approximated as follows,
\begin{equation}
\label{CSD2}
A  = \frac{\tilde e}{\sqrt{2}}
\left(
\begin{array}{r}
0 \\
1 \\
1
\end{array}
\right),
\ \
B = \frac{\tilde b}{\sqrt{3}}
\left(
\begin{array}{r}
1 \\
1 \\
-1
\end{array}
\right),\ \
C \approx  \frac{\tilde c'}{\sqrt{6}}
\left(
\begin{array}{r}
0 \\
0 \\
1
\end{array}
\right),
\end{equation}
since the Dirac neutrino mass matrix is given by $M_{LR} = (A,B,C)$ and in hierarchical
unified
models the third column is dominated by the large 3-3 Yukawa coupling, compared to
which the other
elements in the third column are approximately negligible. In the limit $m_1 \rightarrow 0$
the entire third column $C$ plays no role and may be neglected,
so the fact that $C$ does not satisfy the FD conditions is irrelevant, and in
this limit the CSD conditions for $A,B$ above lead precisely to the TB mixing angles. In
particular we note that in this
limit of CSD there are no NLO corrections to the TB neutrino mixing angles. However, in
practice,
the large 3-3 Yukawa coupling may be expected to lead to a non-zero $m_1$, and in this
case the TB
mixing angles would be expected to be subject to NNLO corrections.
CSD is therefore an example where FD is only violated by soft NNLO corrections.
The neutrino mass $m_1$ can be written
approximately at CSD, in terms of the 3-3 Yukawa coupling, as \footnote{Another formula
for the
mass $m_1$ is derived in Appendix \ref{Amass}.}
\be
m_1 \approx \frac{|c'|^2}{6 X'}.
\ee

Assuming a strong neutrino mass hierarchy $|m_1|\ll |m_2| < |m_3|$, TBR mixing in
Eq.~(\ref{MNS3})
results from a simple modification to CSD, corresponding to allowing a non-zero 1-1
element of the Dirac
neutrino mass matrix,
\begin{equation}
\label{PCSD}
A  = \frac{\tilde e}{\sqrt{2}}
\left(
\begin{array}{c}
re^{-i\delta } \\
1 \\
1
\end{array}
\right), \ \
B = \frac{ \tilde b}{\sqrt{3}}
\left(
\begin{array}{r}
1 \\
1 \\
-1
\end{array}
\right),\ \
C \approx  \frac{ \tilde c'}{\sqrt{6}}
\left(
\begin{array}{r}
0 \\
0 \\
1
\end{array}
\right),
\end{equation}
where we have written $A= \tilde e \Phi_3$, where $\Phi_3$ is the third column of the TBR
matrix in
Eq.~(\ref{MNS3}). This is referred to as Partially Constrained Sequential Dominance
(PCSD) \cite{King:2009qh},
since one of the conditions of CSD is maintained, while another one is violated.
We emphasise that $B$ is unchanged from the case of CSD,
and in particular $B$ is not proportional to the second column of the TBR matrix in
Eq.~(\ref{MNS3}),
and thus, as noted in \cite{King:2009qh}, PCSD violates FD since $A$ and $B$ are not
orthogonal
if $\theta_{13}$ is non-zero.
Thus, one might expect that the TBR form of mixing matrix in Eq.~(\ref{MNS3}) will result
only to LO
with NLO corrections of order $|m_2|/|m_3|$ generally expected.
However the unitarity of the columns $A,B$ which, if exact, would imply no NLO corrections,
is only spoiled by the first element of $A$ proportional to the reactor angle.
Therefore we expect the NLO corrections to PCSD to be suppressed by $\theta_{13}$,
and our numerical results confirm this.

We note that the special cases of SD considered here, namely CSD and PCSD, have a
certain
theoretical elegance since the columns of the Dirac neutrino mass matrix $A,B$ take very
simple forms in
these cases. As discussed in the Introduction these simple forms may be achieved in flavour models based on non-Abelian
discrete family
symmetry (containing triplet representations) and vacuum alignment
of the flavons which break the family symmetry. In particular the CSD and PCSD
approaches
considered here correspond to flavour models in which the neutrino flavour symmetry
inherent in TB and TBR
mixing is achieved in an {\it indirect} way \cite{King:2009ap} from the family symmetry as an accidental
symmetry. In such indirect models \cite{A4-LQ:King,Mohapatra:2003tw,King:2009mk,Z7Z3-LQ,delta27-LQ:King,SO(3)-LQ:King,SU(3)-LQ:Ross}
the flavon vacuum alignment responsible for the simple forms
of the columns $A,B$ of the Dirac neutrino mass matrix arises in supersymmetric models
from the so called D-term vacuum alignment mechanism \cite{Howl:2009ds}.

In general, the MNS matrix may not be as simple as either the TB or the TBR patterns
suggest and thus the special cases of SD such as CSD or PCSD may be regarded as being too restrictive.
Indeed it is quite possible that the actual lepton mixing angles deviate significantly
from those predicted by these special cases, and thus the hierarchical neutrinos satisfy more
general SD conditions (specified more precisely in the next section) than the CSD or PCSD ones.
In such cases it is useful to have analytic expressions for the neutrino mixing angles for the
general case of SD, including the NLO and NNLO corrections, due to arbitrary violations of FD in which the
columns $A,B,C$ do not satisfy complex orthogonality, not even approximately, and these will be
given in subsequent sections.

%%%%%%%%%%%%%%%%%%%%%%%%%%%%%%%%%%%%%%%%%%%%
%%%%
\section{Neutrino mixing angles in general SD to LO}\label{SD}

We consider the case of see-saw mechanism with general SD, involving a
right-handed neutrino mass matrix $M_{RR}$ and a Dirac neutrino matrix $M_{LR}$. We
take $M_{RR}$ to have a diagonal form with real eigenvalues as follows,
\be \label{MRRT}
M_{RR} = \left(\begin{array}{ccc} Y&0&0\\ 0&X&0\\ 0&0&X' \end{array}
\right).
\ee
We also write the complex Dirac neutrino mass matrix $M_{LR}$ in terms
of the general (unconstrained) mass matrix elements $a,b,c,d,e,f,a',b',c'$ as,
\begin{equation}
\label{abc}
M_{LR} = (A,B,C)= \left(\begin{array}{ccc} d&a&a' \\ e&b&b' \\ f&c&c' \end{array} \right).
\end{equation}
In this paper, motivated by unified models where the 3-3 element of the hierarchical mass
matrix is
very large,
we take the two complex elements $a', b'$ to be negligible in comparison to $c'$. The
effective
neutrino mass
matrix can be derived using the see-saw formula \cite{seesaw},
\be
m_{LL} = M_{LR} M^{-1}_{RR} M^T_{LR} = \frac{AA^{T}}{Y} + \frac{BB^{T}}{X} +
\frac{CC^{T}}{X'},
\ee
which gives the following complex symmetric matrix,
\be \label{ssaw}
m_{LL}=\left(\begin{array}{ccc}\frac{a^2}{X}+\frac{d^2}{Y}&\frac{a b}{X}+\frac{d e}{Y}&
\frac{a c}
{X}+\frac{d f}{Y} \\
\frac{a b}{X}+\frac{d e}{Y}&\frac{b^2}{X}+\frac{e^2}{Y}&\frac{b c}{X}+\frac{e f}{Y} \\
\frac{a c}{X}+\frac{d f}{Y} &\frac{b c}{X}+\frac{e f}{Y} &\frac{c'^2}{X'}+\frac{c^2}{X}+\frac{f^2}
{Y}
 \end{array} \right).
\ee
This mass matrix can be diagonalised, for the case of hierarchical neutrino mass matrix, as
presented in Appendix \ref{Diagon}.

Sequential dominance corresponds to the following condition, ignoring $d$ to start with,
\begin{equation}\label{y}
|m_3|=O\left( \frac{|e^2|, |f^2|, |ef|}{Y}\right) \gg |m_2|=O\left( \frac{|xy|}{X}\right)  \gg
|m_1|=O\left( \frac{|{c'}|^2}{X'}\right)
\end{equation}
where $x, y \in a, b, c $ and all Dirac mass matrix elements are
assumed to be complex.
Diagonalising the neutrino mass matrix, the mixing angles can be derived to leading order
(LO), in
the
framework of the seesaw mechanism, as presented in \cite{King:2002nf, King:2002qh}
\footnote{Throughout this paper, for brevity we often write $c_{ij} \equiv \cos (\theta_{ij}),~
s_{ij}
\equiv \sin
(\theta_{ij}),~ t_{ij} \equiv \tan(\theta_{ij})$},
\bea
t_{23} &\approx& \frac{|e|}{|f|} \equiv t_{23}^0 ,\label{theta230}\\
t_{12} &\approx& \frac{|a|}{c_{23} |b| \cos (\phi'_b)- s_{23}  |c| \cos (\phi'_c)} \equiv t_{12}^0,
\label{theta120}\\
\theta_{13} &\approx& e^{i (\phi_2 + \phi_a - \phi_e)}\frac{|a| (e^*b + f^* c)}{(|e|^2+|f|
^2)^{3/2}}
\frac{Y}{X}\equiv \theta_{13}^0, \label{theta130}\
\eea
where some of the Dirac masses were written as $x= |x| e^{i \phi_x}$. The phases $\chi$
and $\phi_2$ are fixed, to give real $\theta_{12}$ and $\theta_{13}$ angles, by:
\bea
c_{23} |b| \sin(\phi'_b) &\approx& s_{23} |c| \sin(\phi'_{c}),\\
\phi_2 &\approx& \phi_e -\phi_a -\phi^*,
\eea
where
\bea
\phi'_b &\equiv& \phi_b - \phi_a -\phi_2-\chi,\\
\phi'_c &\equiv&  \phi_c -\phi_a +\phi_e -\phi_f -\phi_2 -\chi,\\
\phi^* &=& arg(e^*b + f^*c).
\eea

In the large $d$ limit, the angle $\theta_{13}$ can be expressed as follows
\cite{King:2002nf}:
\be \label{theta1330}
\theta_{13} \approx \frac{|d|}{\sqrt{|e|^2 + |f|^2}}\equiv \theta_{13}^0.
\ee
Note that $\theta_{13}$ and $\theta_{13}^0$ are given differently in the small $d$ and large
$d$ cases so we must be careful to distinguish the two limiting cases.
The phases $\phi_2$ and $\phi_3$ appearing in Eq.~(\ref{nmm}) are fixed by:
\bea
\phi_2 &=& \phi_e - \phi_d\\
\phi_3 &=& \phi_f - \phi_d.
\eea

As a simple application of these results, it is clear that
TB neutrino mixing ($\tan \theta_{23} =1$, $\tan
\theta_{12} =1/ \sqrt{2}$ and $\theta_{13}= 0 $) can be achieved by considering CSD
which corresponds to the following set of conditions for the Yukawa couplings \cite{King:2005bj},
\begin{eqnarray}
\label{CSD}
|a|&=&|b|=|c|,\\
|d|&=&0,\\
|e|&=&|f|,\\
e^*b + f^* c&=&0.
\end{eqnarray}
These conditions are also in accordance with Eq.~(\ref{CSD2}).

The results in this section were derived in \cite{King:2002nf, King:2002qh}
at LO, neglecting the NLO and NNLO corrections.
For the remainder of the
paper, we will refer to the LO mixing angles as $t^0_{23}, t^0_{12}$ and $\theta^0_{13}$
respectively,
as indicated in Eqs.~(\ref{theta230}), (\ref{theta120}),
(\ref{theta130}) or (\ref{theta1330}).
Note that the LO atmospheric and solar neutrino mixing angles do not depend on the neutrino
masses, while the reactor angle (in the small $d$ case) does.
However there are expected to be mass dependent corrections to all angles at order NLO
and
NNLO.
As mentioned, near to the CSD limit, the NLO
corrections cancel, leaving only the NNLO corrections arising from the 3-3 Dirac neutrino mass matrix
element $c'$.
This implies that, providing that the NNLO corrections are not too large (i.e.~$m_1$ is not
too large)
the simple LO approximations to the neutrino mixing angles presented in this section are
more accurate than might be expected in the case of CSD.
More generally the LO results may be sufficient for examples of SD close to the FD limit where NLO and NNLO corrections vanish.
However, for many other cases of SD away from the FD limit, the LO results are not sufficient and the
NLO and NNLO corrections need to be considered. This is done in the next section.

%%%%%%%%%%%%%%%%%%%%%%%%%%%%%%%%%%%%%%%%%%%%
%%%%
\section{Neutrino mixing angles in general SD to NLO and NNLO }\label{cor}

In this section, we derive approximate analytic expressions for neutrino mixing angles in
the case
of neutrino mass hierarchy in general SD including NLO and NNLO corections.
The derivations make use of the diagonalisation procedure outlined in Appendix
\ref{Diagon}.
Note that this procedure relies on the fact that the 1-3 reactor angle is small which enables
the
diagonalization to be performed in three stages: first diagonalize the 2-3 block, then the 1-3
outer block,
and finally the 1-2 block. Such a method cannot take into account subleading corrections
resulting from
changes to the 2-3 mixing angle as a result of 1-3 rotations and therefore NLO corrections
to the 2-3 angle
suppressed by the 1-3 angle are not present in the analytic results.

%%%%%%%%%%%%%%%%%%%%%%%%%%%%%%%%%%%%%%%%%%%%
%%%%
\subsection{The atmospheric angle}

As discussed in Appendix \ref{Diagon}, the diagonalisation of the mass matrix involves
applying
the real rotation $R_{23}$ after re-phasing the matrix. This rotation gives rise to two new
mass
terms $\tilde{m}^\nu_{22}$ and $m'_3$ given by Eqs.~(\ref{m22c}),(\ref{m3pn})
respectively. We
consider the 23 block as presented in Eq.~(\ref{diago23}) which gives rise to an expression
for $
\tan(2 \theta_{23})$ in terms of the lower block masses $m_{23}, m_{22}$ and $m_{33}$.
Using Eq.~(\ref{ssaw}), we can substitute for these masses in
terms of the Yukawa couplings to get the
following expression,
\be \label{theta23}
\tan( 2 \theta_{23}) \approx \frac{2 \frac{e f}{Y}( 1+\epsilon_1) e^{i(-\phi_2-\phi_3)} }
{ \frac{f^2}{Y}
(1+\epsilon_2+\eta_1)e^{i (- 2\phi_3)} - \frac{e^2}{Y} (1+\epsilon_3)e^{i
( -2\phi_2)}},
\ee
where we have introduced new parameters $\epsilon_1 , \epsilon_2, \epsilon_3$ and $
\eta_1$,
which are given as follows,
\be
\epsilon_1 = \frac{\frac{b c}{X}}{\frac{e f}{Y}}, ~~\epsilon_2=\frac{\frac{c^2}{X}}{\frac{f^2}
{Y}},~~
\epsilon_3 = \frac{\frac{b^2}{X}}{\frac{e^2}{Y}}, ~~\eta_1 = \frac{\frac{c'^2}{X'}}{\frac{f^2}{Y}}.
\label{epsilon}
\ee
Note that $\epsilon_i, \eta_i$ are of order $m_2/m_3, m_1/ m_3$ respectively, so that
$\epsilon_i$ parametrize the NLO corrections while $\eta_i$ parametrize the NNLO
corrections.

Introducing the small parameter $\delta$ such that $|f| = |e|(1- \delta)$, we get our final
result for the atmospheric angle in SD:
\be \label{theta23n2}
t_{23} \approx t^0_{23} (1+\real (\gamma)),
\ee
where the complex couplings $e, f$ are written in terms of their absolute values and
phases as $e
= |e| e^{i \phi_e}$, $f = |f| e^{i \phi_f}$, respectively, $t^0_{23} \equiv \tan(\theta_{23})|
_{\epsilon_i=0, \eta_i =0}$ is given by Eq.~(\ref{theta230}) and the complex parameter $
\gamma$ is
written as:
\be \label{gamma}
\gamma \approx \frac{1}{2} (\epsilon_3 -\epsilon_2 -\eta_1)
+ \frac{\delta }{2} (\epsilon_3 +  \epsilon_2 -2\epsilon_1 + \eta_1  ).
\ee
%%%%%%%%%%%%%%%%%%%%%%%%%%%%%%%%%%%%%%%%%%
%%%%%%%%%%%%%%%%%%%%%%%%%%%%%%%%%%%%%%%%%%%%
%%%%
\subsection{The reactor angle}
We apply the $R_{13}$ rotation, as outlined in the Appendix, in order to modify the outer
block of
the mass matrix. We consider the reduced matrix that only involves the 13 elements and
this gives
rise to two zeros in the $13, 31$ positions as presented in Eq.~(\ref{Rot13}). The neutrino
angle $
\theta_{13}$ can then be written as,
\be \label{theta13}
\theta_{13} \approx \frac{1}{m'^0_3}( \tilde{m}_{13}^0 (1 - \gamma (s_{23}^0)^2) + e^{-i
\phi_2}
\gamma s_{23}^0(\frac{a b}{X}+ \frac{d e}{Y})) (1- \beta),
\ee
where the masses $m'^0_{3}, m^0_{13},$ are given by Eqs.~(\ref{m3pz}), (\ref{m13z})
respectively.
The complex parameter $\beta$ is given by:
\be \label{beta}
\beta \approx (s^0_{23})^2 \epsilon_3+(c^0_{23})^2 (\epsilon_2+\eta_1) -\epsilon_4  e^{-2 i
\phi_e},
\ee
where the NLO correction parameter $\epsilon_4$ is defined as,
\[ \epsilon_4 = \frac{(b c^0_{23}- c s^0_{23} e^{i (\phi_e - \phi_f)})^2}{X}\left( \frac{|e|^2+|f|^2}
{Y}
\right)^{-1}.\] We can simplify Eq.~(\ref{theta13}) further, after expressing the masses in
terms of the
complex
couplings, by considering two different limits, namelyt the large $d$ limit and the small $d$
limit, as
follows:
\begin{itemize}
\item
{\bf In the large d limit}, $\frac{|d e|}{Y},\frac{|d f|}{Y} >> \frac{|a b|}{X} ,\frac{|a c|}{X}$, the
angle $
\theta_{13}$ can be expressed as,
\be \label{t13dlarge}
\theta_{13} \approx \theta^0_{13}(1- \real( \beta)),
\ee
where the angle $\theta^0 _{13}\equiv \theta_{13}|_{\eta_i= 0,\epsilon_i =0}$ is given by
Eq.~(\ref{theta1330}).
\item
{\bf In the small $d$ limit}, $\frac{|d e|}{Y},\frac{|d f|}{Y}<< \frac{|a b|}{X},\frac{|a c|}{X} $,
which is
usually
the case in CSD,  $\theta_{13}$ can be expressed as,
\bea \label{t13}
\theta_{13}& \approx& \theta^0_{13} \left(1 -\real( \gamma) (s^0_{23})^2 - \real( \beta) \right)
\nonumber \\ &+&
s^0_{23} |\epsilon_5| \left( ( \real \gamma \cos (\phi') -\imag \gamma \sin(\phi'))^2+( \real
\gamma
\sin(\phi')+
\imag \gamma \cos(\phi') )^2\right)^{\frac{1}{2}},
\eea
where $\phi' = \phi_2-2 \phi_e$ and $\theta^0_{13}$, in this limit, is derived in
\cite{King:2002qh}
and given by Eq.~(\ref{theta130}) and the NLO correction parameter $\epsilon_5$ is
defined as,
\be \label{ep5}
\epsilon_5 = \frac{a b}{X}\left( \frac{|e|^2 + |f|^2}{Y}\right)^{-1}.
\ee
\item
{\bf In the PCSD case with non-zero $d$}, we can write the leading result for $\theta_{
13}$ as,
\bea \label{t13PCSD}
\theta_{13}& \approx& \left(\theta^0_{13}+\frac{|d|}{\sqrt{|e|^2+|f|^2}}\right) \left(1 -
\real( \gamma)
(s^0_{23})^2 -
\real( \beta) \right) + s^0_{23} \real (\gamma) \frac{|d||e|}{|e|^2+|f|^2}\nonumber \\ &+&
s^0_{23} |\epsilon_5| \left( ( \real \gamma \cos (\phi') -\imag \gamma \sin(\phi'))^2+( \real
\gamma
\sin(\phi')+
\imag \gamma \cos(\phi') )^2\right)^{\frac{1}{2}},~~~
\eea
where $\theta^0_{13}$ and the parameter $\epsilon_5$ are given by Eqs.~(\ref{theta130}),
(\ref{ep5})
respectively.
\end{itemize}
%%%%%%%%%%%%%%%%%%%%%%%%%%%%%%%%%%%%%%%%%%%%
%%%%
%%%%%%%%%%%%%%%%%%%%%%%%%%%%%%%%%%%%%%%%%%%%
%%%
\subsection{The solar angle}\label{theta12}

As shown in Eq.~(\ref{P1}), applying the phase matrix $P_1$ introduces a new phase $\chi
$ to the
mass matrix. We can then apply the real rotation $R_{12}$, as presented in Eq.~(\ref{R12}),
which
modifies the matrix by putting zeros in the $12, 21$ positions.
Using Eqs.~(\ref{m22c}), (\ref{m12c}) and (\ref{m11c}), we get the following expression for $
\tan(2 \theta_{12})
$,
\be \label{tan212}
\tan(2 \theta_{12})
\approx\frac{2 A B}{B^2 - A^2} \left(1-\gamma (s^0_{23})^2-\zeta_1-\zeta_2  \left(1-\zeta_1 -
\gamma (s^0_{23})^2 \right) \right),
\ee
where, similarly to \cite{King:2002nf}, $A, B$ are expressed in terms of the complex
Yukawa
couplings as,
\be \label{A}
A =\frac{a}{\sqrt{X}},
\ee
\be \label{B}
B = e^{-i(\phi_2 - \chi)}\frac{c^0_{23} b - s^0_{23} c e^{i(\phi_e -\phi_f)}}{\sqrt{X}},
\ee
and the new parameters $\zeta_1$ and $\zeta_2$ are given, in the small $d$ limit, to first
order in
$\gamma$ and $\beta$ as,
\bea
\zeta_1 &\approx& e^{-i (\phi_3+\chi)}\left( \frac{acs^0_{23}}{ A B X} \right) \gamma
\label{zeta1},\\
\zeta_2  &\approx& \frac{1}{B^2- A^2} \left(\eta_2 \left(\frac{b^2}{X}+\frac{e^2}{Y}\right) e^{- 2
i \chi} -
B^2\beta \right) \label{zeta2},
\eea
where $\eta_2$ is given by,
\be \label{eta2}
\eta_2 =\frac{c'^2}{X'}\left( \frac{|e|^2+|f|^2}{Y}\right)^{-1}.
\ee

Using Eq.~(\ref{tan212}) we obtain our final result for the solar mixing
angle in SD:
\be  \label{t12}
t_{12} \approx  t^0_{12}(1- \real (\zeta')).
\ee
where $t^0_{12} \equiv  \tan(\theta_{12})|_{\eta_i=0, \epsilon_i =0}$ is given by
Eq.~(\ref{theta120}).
The new parameter $\zeta'$ is given as,
\be \label{zetap}
 \zeta' \approx \frac{B^2 -A^2}{B^2 +A^2} \left(\gamma (s^0_{23})^2+ \zeta_1 +\zeta_2 (1-
\zeta_1-
\gamma (s^0_{23})^2) \right).
 \ee
%%%%%%%%%%%%%%%%%%%%%%%%%%%%%%%%%%%%%%%%%%%%
%%%
%%%%%%%%%%%%%%%%%%%%%%%%%%%%%%%%%%%%%%%%%%%%
%%%%
\section{Neutrino mixing angles in special cases of SD} \label{pcsd}
In this section, we will look at how the rather complicated analytic results for the mixing
angles in
SD, derived in
the
previous section to NLO and NNLO, can be simplified in the special cases of SD
corresponding
to CSD and PCSD. For simplicity, we shall take $\phi_e = \pi$ and all the remaining
Yukawa
phases are taken
to be zero
except $\phi_{c'}$ which is left general.

\subsection{CSD}
CSD corresponds to SD with the constraints defined in Eq.~(\ref{CSD}).
A particular example of CSD was discussed in Eq.~(\ref{CSD2}) below which
we presented an argument for why we expect the NLO corrections to vanish for CSD
leaving only the NNLO corrections dependent on $m_1$ whose magnitude
is governed by the 3-3 element of the Dirac neutrino mass matrix $c'$.
Using the analytic results for neutrino mixing angles in SD derived to NLO and NNLO
we can explicitly verify that the NLO corrections vanish for CSD.

\subsubsection{The atmospheric angle}
The atmospheric angle in Eq.~(\ref{theta23n2}) becomes, in CSD,
\be \label{theta23n20}
t^{CSD}_{23} \approx 1+\real (\gamma^{CSD}),
\ee
which involves a correction $\gamma$ in Eq.~(\ref{gamma}) which depends on the NLO
parameters
$\epsilon_i$ and the NNLO parameters $\eta_i$ in Eq.~(\ref{epsilon}).
In CSD the conditions in Eq.~(\ref{CSD}) imply that the $\epsilon_i$ are all equal and $
\delta =0$.
From Eq.~(\ref{gamma}) it is clear that the NLO contributions to $\gamma$ described by the
$
\epsilon_i$ cancel.
This implies that the atmospheric angle is corrected by $\gamma$ which only involves
NNLO
corrections given
by,
\be \label{gCSD0}
\gamma^{CSD}  \approx -\frac{\eta_1}{2} \approx  -\frac{1}{2} \frac{|c'|^2 Y}{ |e|^2X'} e^{i 2
\phi_{c'}}.
\ee

\subsubsection{The reactor angle}
Turning to the reactor angle $\theta_{13}$, we only need to consider the expression valid in
the
small $d$ limit
given by
Eq.~(\ref{t13}). Imposing the CSD conditions in Eq.~(\ref{CSD}),
the angle $\theta^0_{13}$ becomes exactly zero as
can seen from Eq.~(\ref{theta130}). As a result, the first term of Eq.~(\ref{t13}) vanishes.
Also the third term vanishes for CSD. We are
only left with the second term of order $\eta \epsilon$,
\be
\theta_{13}^{CSD} \approx s_{23}^0\epsilon_5Re(\gamma^{CSD}) =
\frac{1}{ 4 \sqrt{2}} \frac{|b|^2 Y^2 |c'|^2}{|e|^4 X X'} \cos(2 \phi_{c'}).
\ee
This implies that the reactor angle is given by a term proportional to NNLO~$\cdot$~NLO
corrections.

\subsubsection{The solar angle}
The solar angle in Eq.~(\ref{t12}) becomes, in CSD,
\be  \label{t12C}
t_{12}^{CSD} \approx  \frac{1}{\sqrt{2}}(1- \real (\zeta'^{CSD}))\:,
\ee
which involves a correction $\zeta'$ in Eq.~(\ref{zetap}) which we approximate here to,
\be \label{zetap0}
 \zeta'^{CSD} \approx \frac{1}{3}\left( \frac{\gamma^{CSD}}{2} + \zeta_1^{CSD} +
\zeta_2^{CSD}
\right).
 \ee
which depends on $\gamma^{CSD}$ as well as
the parameters $\zeta_1,\zeta_2$ in Eqs.~(\ref{zeta1}), (\ref{zeta2}),
which also depend on $\beta$ in Eq.~(\ref{beta}).
The parameter $\gamma^{CSD}$ takes the simplified form above in Eq.~(\ref{gCSD0}).
The parameters $\beta, \zeta_1, \zeta_2$, given by
Eqs.~(\ref{beta}), (\ref{zeta1}), (\ref{zeta2}),
can also be
simplified in CSD as,
\bea
\beta^{CSD} &\approx&  - \gamma^{CSD},\\
\zeta_1^{CSD} &\approx&  - \frac{\gamma^{CSD}}{2},\\
\zeta_2^{CSD} &\approx& \frac{1}{2} \frac{|c'|^2 X}{ |a|^2X'} e^{i 2 \phi_{c'}}.
\eea
The important point to note is that the solar angle is corrected only by NNLO corrections.
Indeed all corrections to neutrino mixing angles in CSD vanish at NLO, and first occur only
at
NNLO, as anticipated.

We have compared the analytical results in CSD with the numerical results calculated
using the MPT/REAP software package
\cite{Antusch:2005gp} and found good agreement. The numerical results are shown in
Fig.~\ref{fig:plots} (left panel). While $\theta_{13}$ is rather stable
and remains close to zero, the plot in Fig.~\ref{fig:plots} illustrates that the deviation of $
\theta_{12}$ from the tri-bimaximal value is about twice the
deviation of $\theta_{23}$ from $\pi/4$. It also illustrates the above-derived dependence of
the results on the phase $\phi_{c'}$.

\subsection{PCSD}

PCSD is similar to CSD defined in Eq.~(\ref{CSD}), but with a non-zero value of $d$.
A particular example of PCSD was discussed in Eq.~(\ref{PCSD}).
In this example the NLO corrections do not vanish but are suppressed by the reactor angle.
However, in deriving the analytic results for neutrino mixing angles in SD to NLO and
NNLO,
such corrections are beyond the scope of our approximations. However the results below
have the
simple property that, as in the case of CSD, they only depend on NNLO corrections.
This demonstrates that the NLO corrections vanish for PCSD, neglecting
NLO terms suppressed by the reactor angle.

\subsubsection{The atmospheric angle}
The atmospheric angle in Eq.~(\ref{theta23n2}) becomes, in PCSD,
\be \label{theta23n200}
t^{PCSD}_{23} \approx 1+\real (\gamma^{PCSD}),
\ee
where it is easy to see that the result is identical to the case of CSD,
\be \label{gCSD00}
\gamma^{PCSD} = \gamma^{CSD}.
\ee
This implies that the atmospheric angle correction only involves NNLO corrections, as in
the case
of CSD.

\subsubsection{The reactor angle}
Turning to the reactor angle $\theta_{13}$, we find for PCSD,
using Eq.~(\ref{theta13}),
\be
\theta_{13}^{PCSD} \approx \theta_{13}^0(1+\real(\gamma^{PCSD} ))-\frac{\real
(\gamma^{PCSD})}{2}\frac{|b|
^2Y}{\sqrt{2}|
e|^2X},
\ee
where the LO expression is given by the large $d$ approximation in Eq.~(\ref{theta1330}),
\be
\theta_{13}^0 \approx \frac{|d|}{\sqrt{2}|e|}.
\label{theta130PSD}
\ee
The reactor angle therefore only receives NNLO corrections proportional to $
\gamma^{PCSD}=
\gamma^{CSD}
$.

\subsubsection{The solar angle}
The solar angle in Eq.~(\ref{t12}) becomes, in PCSD,
\be  \label{t1200}
t_{12}^{PCSD} \approx  \frac{1}{\sqrt{2}}(1- \real (\zeta'^{PCSD}))\:,
\ee
which involves a correction $\zeta'$ in Eq.~(\ref{zetap}) which we approximate here to,
\be \label{zetap0C}
 \zeta'^{PCSD} \approx \frac{1}{3}\left( \frac{\gamma^{PCSD}}{2} + \zeta_1^{PCSD} +
\zeta_2^{PCSD} \right)\:,
 \ee
where
\bea
\zeta_1^{PCSD} &\approx&  \gamma^{PCSD}(1+ \theta_{13}^0\frac{\sqrt{2}|e|^2X}{|b|^2Y})
\:,\\
\zeta_2^{PCSD} &\approx& \zeta_2^{CSD}+\sqrt{2}\theta_{13}^0\gamma^{PCSD}+
(\theta_{13}^0)^2\:,
\eea
and where $\theta_{13}^0$ is given in Eq.~(\ref{theta130PSD}.
The important point to note is that the solar angle has no NLO corrections.
However there is a correction of order $(\theta_{13}^0)^2$.
This can be understood if one thinks of the diagonalization procedure
since the 2-3 diagonalization in PCSD is the same as in CSD (since the 2-3 blocks are
identical),
while the subsequent 1-3 diagonalizaton will correct the 1-1 element at order $
(\theta_{13}^0)^2$,
and this will lead to the above correction to the solar angle at order $(\theta_{13}^0)^2$.
However the results show that the corrections to neutrino mixing angles in PCSD vanish at
NLO,
neglecting NLO terms suppressed by the reactor angle,
with only the NNLO corrections remaining to this approximation.

\begin{figure}
\centering
\includegraphics[scale=0.67]{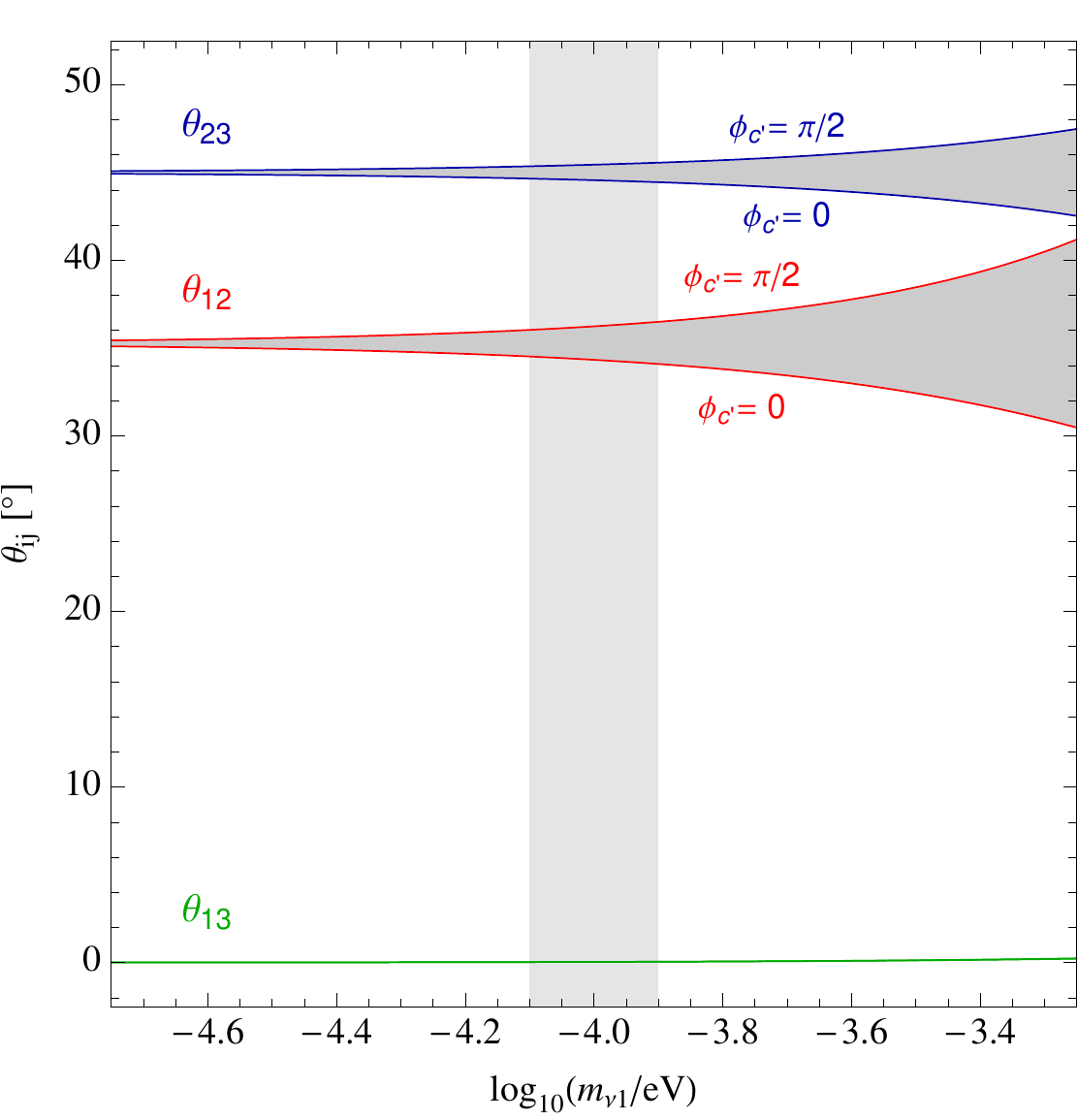}\quad
\includegraphics[scale=0.67]{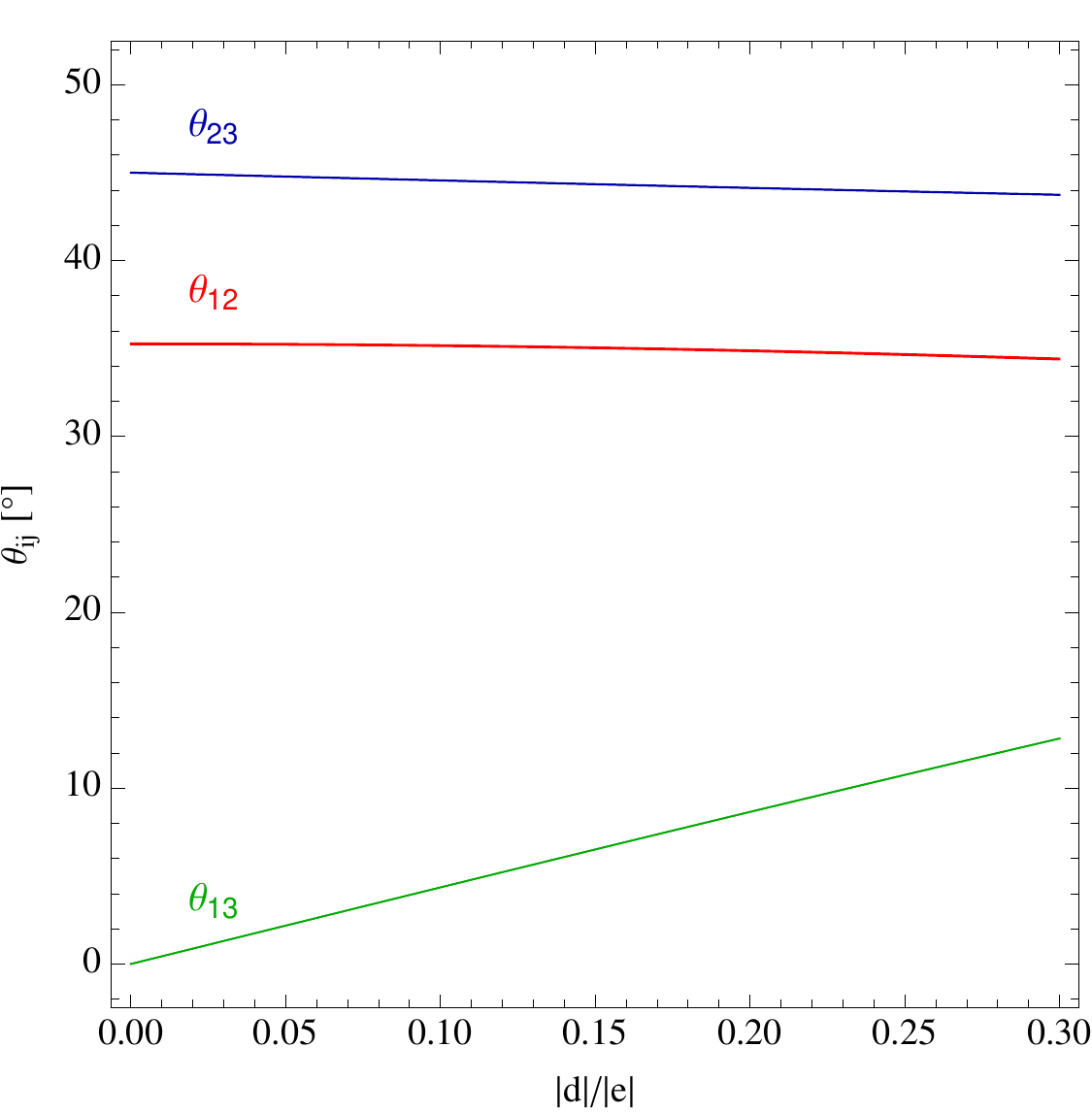}
\caption{Numerical results using the MPT/REAP package \cite{Antusch:2005gp}
for the corrections to the lepton mixing angles in CSD (left panel) and PCSD (right panel),
where in the latter
case $c'$ has been set to zero. In CSD (left panel), the NNLO correction depends on the
mass of the lightest neutrino,
$m_{\nu_1}$, and the light-grey vertical band shows the typical expectation for
$m_{\nu_1}$ in some classes of
GUT models. While $\theta_{13}$ is rather stable and remains close to zero, the deviation
of $\theta_{12}$ from
the tri-bimaximal value is about twice the deviation of $\theta_{23}$ from $\pi/4$. The two
extreme cases
with $\phi_{c'} = 0, \pi/2$ (note that the change of both, $\theta_{12}$ and $\theta_{23}$ is
approximately
proportional to $\cos (2 \phi_{c'})$) are displayed and the grey area between the curves
corresponds to other
values of $\phi_{c'}$. In PCSD (right panel) with zero $c'$, $\theta_{12}$ and $\theta_{23}$
are nearly unchanged and
only $\theta_{13}$ increases with increasing $|d|/|e| = |d|/|f|$. The results agree well with
the analytic results,
and confirm that in the case of PCSD there there are small NLO corrections to $
\theta_{23}$ and $\theta_{12}$ proportional to $\theta_{13}$ which are not
seen in the analytic results. \label{fig:plots}}
\end{figure}

The analytical results in PCSD agree well with the numerical results calculated using the
MPT/REAP software package
\cite{Antusch:2005gp}. For the special case of zero $c'$ the numerical results are displayed
in Fig.~\ref{fig:plots} (right panel). In accordance with the
analytical results, $\theta_{12}$ and $\theta_{23}$ are nearly unchanged and
only $\theta_{13}$ increases with increasing $|d|/|e| = |d|/|f|$.

%%%%%%%%%%%%%%%%%%%%%%%%%%%%%%%%%%%%%%%%%%%%
%%%%
%%%%%%%%%%%%%%%%%%%%%%%%%%%%%%%%%%%%%%%%%%%%
%%%%
\section{Analytical vs.\ numerical results in two example models} \label{numerical}
In the previous section, analytic expressions of the mixing angles, involving NLO and
NNLO
corrections, were derived. Approximate results in the case of CSD and PCSD were also
presented
and the NLO corrections vanished in both cases. In this section, we evaluate the analytic
results for
two different numerical
GUT inspired models. These are
models of light
sequential dominance (LSD) and heavy sequential dominance (HSD)
\cite{King:1998jw, King:2004} previously studied in \cite{Boudjemaa:2008jf}. We also
present
numerical results for the neutrino masses, presented in Appendix \ref{Amass}, as well as
the difference in neutrino masses squared for both models.

%%%%%%%%%%%%%%%%%%%%%%%%%%%%%%%%%%%%%%%%%%%%
%%%%
\subsection{Results for the LSD model}
%%%%%%%%%%%%%%%%%%%%%%%%%%%%%%%%%%%%%%%%%%%%
%%%%
\subsubsection{The CSD case} \label{csd}

We consider a simple numerical example
with a diagonal right-handed neutrino Majorana mass matrix $M_{RR}$ given by,
\be \label{LSM}
M_{RR}=\left( \begin{array}{ccc} 5.1 \times 10^{-9} & 0 & 0 \\
0 & 7.05 \times 10^{-9} & 0 \\
0 & 0 &1 \end{array} \right) M_3,
\ee
where $M_3 = 10^{16} GeV$. We also consider a neutrino Yukawa matrix of the form,
\be \label{Yn}
Y^\nu_{LR}=\left( \begin{array}{ccc} 0 & a & 0 \\
e & b  & 0 \\
f & c & c' \end{array} \right),
\ee
where we have taken the complex coupling $d$ to be zero. As required by CSD, we take $|
a| = |b|
=|c| = 8.125
\times 10^{-5}$ and $|e| =|f| =2.125 \times 10^{-4}$. The value of the third family coupling
$c'$ is
taken to be $|
c'| = 0.5809$. We choose all the phases of the Yukawa couplings to be zero except $\phi_e
$ ($
\phi_e = \pi$).
 \begin{table}[hbtp]

    \centering

    \begin{tabular}{|l|c|c|c|c|c|c|c|c|}

    \hline

   Parameter &$|d|$ &$|c'|$& $\theta_{23}
( \,^{\circ})$ & $\theta_{13}( \,^{\circ})$ & $\theta_{12}( \,^{\circ})$ &$m_1$ (eV)&$m_2$
(eV)&$m_3$
(eV)\\

    \hline

    Analytic &$0$ &$0.5809$& 44.44 & 0.04 & 33.75&$1.52 \times 10^{-4}$&$8.8 \times
10^{-3}$&
$0.055$\\

        \hline

    MPT/REAP &$0$&$0.5809$& 44.38 & 0.05 & 33.69& $1.61 \times 10^{-4}$& $8.8 \times
10^{-3}$&$
0.054$\\

    \hline

    \end{tabular}

    \caption{ Analytical vs. numerical results for the mixing angles and masses, evaluated in the CSD case
with $c' \not= 0$, for a model with light sequential dominance.}

    \label{tabn1}

\end{table}

Numerical results for the mixing angles, evaluated in CSD using the analytic formulae, are
presented in
Tab.~\ref{tabn1}. This table also shows numerical results obtained using MPT/REAP
package
\cite{Antusch:2005gp}, which appear to be
very close to the ones obtained through the analytic approach. We note that here and in the
remainder of
the paper, the MPT/REAP results were
evaluated using the MPT package without considering RG running. As can be seen from
Tab.~\ref{tabn1}, all the values of
the mixing angles are slightly deviated from their TB values and this is mainly due to the
presence
of the non-zero 3-3 Yukawa coupling $c'$.\footnote{In the limit $c' = 0$, the analytic results
give
exact TB
values ($\theta_{23}=45.00^\circ,~\theta_{12}=35.26^\circ,~\theta_{13}=0.00^\circ$).}
In addition, we present numerical results for the neutrino masses $m_1, m_2$ and $m_3$
given by Eqs.~(\ref{m1}), (\ref{m2}), (\ref{m3}), using both MPT/REAP and the analytic
formulae. As
presented in Tab.~\ref{tabn1}, we can see that the MPT/REAP and the analytic results are
very close
particularly in the case of $m_2$.

%%%%%%%%%%%%%%%%%%%%%%%%%%%%%%%%%%%%%%%%%%
\subsubsection{The PCSD case}

We consider the previous LSD model in the case of PCSD with non-zero Yukawa coupling
$d= 0.2
|e| $, $|e| =
2.125 \times
10^{-4}$ and $|c'| =0$. Keeping all the other conditions of CSD
satisfied as outlined in Section \ref{csd}, we found that the numerical values of all the
mixing
angles are deviated from their TB values particularly the reactor angle $\theta_{13}$ which
becomes larger
than zero and takes a value of
$8.22^\circ$ as shown in Tab.~\ref{tabn2}. This large value satisfies the
predictions of TBR mixing and it is in good agreement with the most recent experimental
results
\cite{Fogli:2009ce}.

 MPT/REAP results for the neutrino mixing angles in this case are slightly different than the
analytic
results as
presented in Tab.~\ref{tabn2}. This is mainly due to the approximate nature of the
diagonalisation
procedure that we followed in this paper. Tab.~\ref{tabn2} also shows numerical results for
the
neutrino masses $m_1,
m_2$ and
$m_3$ evaluated in PCSD using both MPT/REAP and the analytic expressions. As
expected, the neutrino
mass
$m_1$ is
exactly zero in this case due to the vanishing  NNLO corrections. We note that the results
for the masses $m_2,
m_3$ in the analytic case are slightly different than the MPT/REAP case. This is also due to
the
difference in the diagonalisation approaches considered in the two cases.

\begin{table}[hbtp]

    \centering

    \begin{tabular}{|l|c|c|c|c|c|c|c|c|}

    \hline

   Parameter &$|d|$ &$|c'|$& $\theta_{23}
( \,^{\circ})$ & $\theta_{13}( \,^{\circ})$ & $\theta_{12}( \,^{\circ})$&$m_1$  (eV)&$m_2$
(eV)&
$m_3$  (eV) \\

    \hline

        Analytic&$0.2 |e|$ &$0$& 45.00 & 8.10 & 35.08&$0$&$8.5 \times 10^{-3}$&$5.38
\times
10^{-2}$\\

    \hline

      MPT/REAP &$0.2 |e|$&$0$& 44.29 & 8.53 & 34.89&0&$8.4 \times 10^{-3}$&$5.4 \times
10^{-2}$\\

    \hline

    \end{tabular}

    \caption{ Analytical vs. numerical results for the neutrino mixing angles and masses, evaluated in the
PCSD
case for a
model with light sequential dominance, with $c' = 0$ and $|d| = 0.2 |e|$.}

    \label{tabn2}

\end{table}

In order to compare our numerical values to experimental data, we present numerical
results for
the difference in the squares of neutrino masses $\Delta m^2_{sol}$ and $\Delta m^2_{atm}
$,
evaluated for the LSD model in CSD and PCSD, as shown in Tab.~\ref{tabn21}. These
results are evaluated at the SD
cases using both the analytic results as well as MPT/REAP. The numerical results, as
shown in
Tab.~\ref{tabn21}, are within the most recent experimental ranges presented in
\cite{Fogli:2009ce} particularly
for the value of $\Delta m^2_{sol}$ in CSD which is very close to the best fit value of $7.6
\times 10^{-5}
eV^2$.

\begin{table}[hbtp]

    \centering

    \begin{tabular}{|l|c|c|c|c|c|}

    \hline

   Parameter & Analytic & MPT/REAP& Analytic & MPT/REAP\\
    \hline
  SD limit & CSD & CSD & PCSD& PCSD\\
    \hline
     $\Delta m^2_{sol}$ ($eV^2$)& $7.5 \times 10^{-5}$ & $7.5 \times 10^{-5}$ & $7.3 \times
10^{-5}$
&$7.1 \times 10^{-5}$\\
    \hline

       $\Delta m^2_{atm}$ ($eV^2$)& $2.11\times 10^{-3}$ & $2.04\times 10^{-3}$ & $2.05
\times
10^{-3}$ & $2.1 \times 10^{-3}$\\

    \hline

    \end{tabular}

    \caption{ Analytical vs. numerical results of the difference in the squares of
neutrino
masses ($\Delta m^2_{sol}$ and $\Delta m^2_{atm}$) evaluated for the LSD model. The
results
are presented at CSD with non-zero $c'$ as well as the PCSD case with zero $c'$ and non-
zero
coupling $|d| = 0.2 |e|$. }
    \label{tabn21}

\end{table}

%%%%%%%%%%%%%%%%%%%%%%%%%%%%%%%%%%%%%%%%%%%%
%%%%
\subsection{Results for the HSD model}

To check the generality of our numerical results, we consider another model with heavy
sequential
dominance (HSD). The right-handed neutrino Majorana mass matrix $M_{RR}$, in this
case, is
given by,
\be
M_{RR}=\left( \begin{array}{ccc} 3.991 \times 10^8 & 0 & 0 \\
0 & 5.8 \times 10^{10} & 0 \\
0 & 0 & 5.021 \times 10^{14} \end{array} \right).
\ee
This model satisfies HSD where the dominant contribution to the neutrino mass is coming
from the
heaviest right-handed neutrino. The neutrino Yukawa matrix is of the form given in
Eq.~(\ref{Yn})
with the following values of the Yukawa couplings: $|a| = |b| = |c| = 2.401 \times 10^{-3}$, $|
e| = |f| =
0.677$ and $|c'| = 2.992 \times 10^{-5}$. Similarly to the LSD model, we take all the phases
of the
Yukawa
couplings to be zero except the coupling $e$ ($\phi_e = \pi$).

Analytic and MPT/REAP results of the mixing angles and masses, in CSD, are presented
and
compared
as shown
in Tab.~\ref{tabn3}. We note that, for this model, the values of the mixing angles are much
closer to their
TB
values compared to the LSD model, which is mainly due to the smallness of the 3-3
Yukawa
coupling $c'$
in this case. We also present results for the PCSD case with non-zero 1-1 Yukawa coupling
($d$), as shown in
Tab.~\ref{tabn4}. Analogously to the LSD model, the reactor angle for the HSD model, is
found to be large and within
the
recent experimental range presented in \cite{Fogli:2009ce}. The neutrino mass $m_1$ is
exactly
zero at the
PCSD case with $c' =0$ as expected.

 \begin{table}[hbtp]

      \centering

        \begin{tabular}{|l|c|c|c|c|c|c|c|c|}

    \hline

   Parameter &$|d|$ &$|c'|$& $\theta_{23}
( \,^{\circ})$ & $\theta_{13}( \,^{\circ})$ & $\theta_{12}( \,^{\circ})$ &$m_1$  (eV)&$m_2$
(eV)&
$m_3$  (eV)\\

    \hline

    Analytic &$0$ &$2.99 \times 10^{-5}$& 44.96 & $0.003$  & 35.18&$1.01 \times 10^{-5}$&
$9
\times 10^{-3}$&
$0.055$\\

        \hline

    MPT/REAP &$0$&$2.99 \times 10^{-5}$& 44.96 & $0.003$  & 35.16& $1.1 \times
10^{-5}$&$ 9
\times 10^{-3}$&
$0.055$\\

    \hline

    \end{tabular}

    \caption{ Analytical vs. numerical results for the neutrino mixing angles and masses, evaluated in CSD
with
$c'
\not=
0$, for
a model
with heavy sequential dominance.}

    \label{tabn3}

\end{table}

 \begin{table}[hbtp]

    \centering

    \begin{tabular}{|l|c|c|c|c|c|c|c|c|}

    \hline

   Parameter &$|d|$ &$|c'|$& $\theta_{23}
( \,^{\circ})$ & $\theta_{13}( \,^{\circ})$ & $\theta_{12}( \,^{\circ})$&$m_1$  (eV)&$m_2$
(eV)&
$m_3$  (eV) \\

    \hline

        Analytic&$0.2 |e|$ &$0$&  45.00& $8.10$  & 35.08&$0$&$9
\times 10^{-3}$&$0.055$\\
    \hline

      MPT/REAP &$0.2 |e|$&$0$&$44.27$& $ 8.55$  & $34.89$&$ 0$&$8.9
\times 10^{-3}$&$0.056$\\

    \hline

    \end{tabular}

    \caption{ Analytical vs. numerical results for the neutrino mixing angles and masses, evaluated in the
PCSD
case
for a model with heavy sequential dominance, with $c' = 0$ and $|d| = 0.2 |e|$.}

    \label{tabn4}

\end{table}

Similarly to the LSD model, we present numerical results for the difference in the squares of
neutrino masses $\Delta m^2_{sol}$ and $\Delta m^2_{atm}$, evaluated for the HSD
model, as
shown in Tab.~\ref{tabn41}. The results for this model, which are also presented at both SD
cases
using analytic results as well as MPT/REAP, are within the most recent experimental
ranges presented in
\cite{Fogli:2009ce}. We note that the values of $\Delta m^2_{sol}$, in all cases presented in
Tab.~\ref{tabn41}, are closer to the
upper limit of the $3 \sigma$ experimental range \cite{Fogli:2009ce}.
\begin{table}[hbtp]

    \centering

    \begin{tabular}{|l|c|c|c|c|c|}

    \hline

   Parameter & Analytic & MPT/REAP& Analytic & MPT/REAP\\
   \hline
  SD limit & CSD & CSD & PCSD& PCSD\\
    \hline

     $\Delta m^2_{sol}$ ($eV^2$)& $8.2 \times 10^{-5}$ & $8.15 \times 10^{-5}$ & $8.2 \times
10^{-5}$ &$8 \times 10^{-5}$\\
    \hline

       $\Delta m^2_{atm}$ ($eV^2$)& $2.16 \times 10^{-3}$ & $2.13\times 10^{-3}$ & $2.16
\times
10^{-3}$ & $2.2 \times 10^{-3}$\\

    \hline

    \end{tabular}

    \caption{ Analytical vs. numerical results of the difference in the squares of
neutrino
masses ($\Delta m^2_{sol}$ and $\Delta m^2_{atm}$) evaluated for the HSD model. The
results
are presented at CSD with non-zero $c'$ as well as the PCSD case with zero $c'$ and non-
zero
coupling $|d| = 0.2 |e|$. }
    \label{tabn41}

\end{table}

To summarize, the comparison between the analytical results and the numerical values
using MPT/REAP has shown small differences, which are however
within the expected range due to the approximate nature of the diagonalisation procedure
followed in this paper. Explicitly, $\theta_{12}$ from the
analytical results has been larger by about $0.2^\circ$ than the MPT/REAP value, $
\theta_{23}$ is larger by $0.7^\circ$ while the reactor angle $
\theta_{13}$ is smaller by about $0.4^\circ$ than the MPT/REAP value for both models. We
have also presented numerical results for the neutrino masses
as well as the difference in the squares of neutrino masses ($\Delta m^2_{sol}$ and $\Delta
m^2_{atm}$), in the SD cases, using MPT/REAP and the
analytic formulae. The numerical results for $\Delta m^2_{sol}$ and $\Delta m^2_{atm}$
are within the most recent experimental ranges  \cite{Fogli:2009ce}.

%%%%%%%%%%%%%%%%%%%%%%%%%%%%%%%%%%%%%%%%%%%%
%%%%
\section{Summary and Conclusions}\label{conc}

SD provides a natural way of having large mixing angles together with hierarchical
neutrino masses
from the type I see-saw mechanism. In this paper we have derived analytic expressions
for the neutrino mixing angles including
the NLO and NNLO corrections arising from the second lightest and lightest neutrino
masses.
We have pointed out that for special cases of SD corresponding to form dominance
the NLO and NNLO corrections both vanish.
This observation helps to explain why the LO results for SD are sometimes
observed to work better than expected in particular cases.
For example, we have studied tri-bimaximal mixing via constrained sequential dominance
which involves only a NNLO correction as an example of softly broken FD
where the corrections vanish in the strong hierarchical limit.
We have also considered tri-bimaximal-reactor
mixing via partially constrained sequential dominance
and shown that this involves NLO corrections to neutrino mixing angles suppressed by the
small reactor angle
which renders such corrections negligible. This supports the PCSD proposal as a robust
way of obtaining
a significant non-zero reactor angle with the atmospheric and solar angles remaining close
to their
tri-bimaximal values to an accuracy of a few per cent.

We have evaluated these analytic results for two explicit GUT inspired models of so-called
LSD
type and HSD type including, in the case of CSD, the deviations from exact
TB mixing due to the presence of a non-zero third family Yukawa coupling. For both models
the
analytic results agree well with the numerical results obtained from extracting the mixing
angles using the MPT
tool provided with MPT/REAP. In the CSD case the absence of NLO corrections is confirmed
numerically
while the soft NNLO corrections are observed.
We have also seen that the analytic expressions in the case of PCSD with zero
third family Yukawa coupling involves no NLO corrections in the approximation that
NLO corrections multiplied by the reactor angle are dropped.
The numerical results for PCSD confirm that the atmospheric and solar angles remain very
close to their TB
values while the reactor angle, $\theta_{13}$, is much larger than zero, and confirm that the
NLO corrections are suppressed by the reactor angle.
If a sizeable reactor angle is measured it may therefore be possible to retain the
tri-bimaximal predictions for the atmospheric and solar angles to a few per
cent accuracy using PCSD. We note, however, that this neglects other model dependent
charged lepton, RG and canonical normalization corrections
which have been considered elsewhere.

To conclude, we have derived approximate analytic results which allow to extract the
neutrino mixing angles in the framework of see-saw mechanism with
sequential dominance including NLO and NNLO corrections from $m_2$ and $m_1$. We
have
shown that, for the special case of FD, the NLO and NNLO corrections vanish. CSD is an
example
where FD is only violated softly leading to NNLO corrections only, a result supported by
numerical evaluation. Such NNLO corrections are typical in many classes of GUT models with hierarchical
Yukawa matrices which lead to LSD with $m_1$ arising from the large
third family Yukawa coupling. We emphasise that in realistic models, FD and CSD are enforced by family symmetry
and so in testing these schemes one is really testing particular models based on family symmetry.
It is possible for FD to yield either a normal mass ordering or an inverted one, while CSD always
corresponds to a normal mass hierarchy.
We have also shown that, in the case of PCSD, the
NLO corrections are suppressed by the reactor angle, and have again verified this result
numerically.
In other special cases of SD which are close to the FD limit, one can similarly expect the
NLO and NNLO corrections
to neutrino mixing angles to be suppressed.
However, if significant deviations from TB mixing are observed, then the more general cases of SD away from the FD
limit, as considered in this paper, would become relevant. The analytic results
for neutrino mixing angles in SD presented here, including the NLO and NNLO corrections,
are therefore applicable to a wider class of models
and may be reliably used when confronting such models with high precision neutrino data,
providing useful insights beyond a purely numerical analysis.

\section*{Acknowledgments}

SFK acknowledges support from the STFC Rolling Grant ST/G000557/1
and is grateful to the Royal Society for a Leverhulme Trust Senior Research Fellowship.
S.A.\  acknowledges partial support from the DFG cluster of excellence ``Origin and
Structure
of the Universe''. S.B.\ would like to thank the Algerian Ministry of Higher Education and
Scientific Research for the support of a scholarship.

%%%%%%%%%%%%%%%%%%%%%%%%%%%%%%%%%%%%%%%%%%%%
%%%%
\appendix
%%%%%%%%%%%%%%%%%%%%%%%%%%%%%%%%%%%%%%%%%%%%
%%%%
\section{Diagonalisation of left-handed neutrino matrix} \label{Diagon}

In this Appendix, we will briefly review the procedure of diagonalising the neutrino mass
matrix
following \cite{King:2002nf} closely. We start by writing the left-handed neutrino mass
matrix as,
\be
m_{LL}^\nu = \left( \begin{array}{ccc}
m_{11} & m_{12} & m_{13} \\
m_{12} & m_{22} & m_{23} \\
m_{13} & m_{23} & m_{33}
\end{array}
\right)  \equiv
\left( \begin{array}{ccc}
|m_{11}|e^{i\phi_{11}}
& |m_{12}|e^{i\phi_{12}}
& |m_{13}|e^{i\phi_{13}}  \\
|m_{12}|e^{i\phi_{12}}
& |m_{22}|e^{i\phi_{22}}
& |m_{23}|e^{i\phi_{23}} \\
|m_{13}|e^{i\phi_{13}}
& |m_{23}|e^{i\phi_{23}}
& |m_{33}|e^{i\phi_{33}}
\end{array}
\right)
\label{nm}
\ee
In general, we diagonalise a complex, hierarchical, neutrino matrix by following a
sequence of
transformations \cite{King:2002nf},
\be \label{diagon}
{P_3^{\nu_L}}^* {R_{12}}^T {P_1^{\nu_L}}^*{R_{13}^{\nu_L}}^T{R_{23}^{\nu_L}}^T
{P_2^{\nu_L}}^*
m_{LL}^\nu P_2^{\nu_L} R_{23}^{\nu_L}  R_{13}^{\nu_L} P_1^{\nu_L} R_{12}^{\nu_L}
P_3^{\nu_L} = \left( \begin{array}{ccc}
m_1&0&0\\
0&m_2&0\\
0&0&m_3
\end{array}\right),
\ee
where the resulting matrix includes the three different neutrino masses $m_1, m_2$ and
$m_3$.
$R_{ij},~i,j=\{1,2,3\}$ are a set of real rotations, involving the Euler angles $\theta_{ij}$,
which can
be written as,
\bea
R_{23} &=& \left(\begin{array}{ccc}
1&0&0\\
0&c_{23}&s_{23}\\
0&-s_{23}&c_{23} \label{R23}\\
\end{array}\right)\\
R_{13} &=& \left(\begin{array}{ccc}
c_{13}&0&s_{13}\\
0&1&0\\
-s_{13}&0&c_{13}\\
\end{array}\right)\\
R_{12} &=& \left(\begin{array}{ccc}
c_{12}&s_{12}&0\\
-s_{12}&c_{12}&0\\
0&0&1\\
\end{array}\right).\\ \nonumber
\eea
The matrices $P_i$ in Eq.~(\ref{diagon}) are the phase matrices, involving the phases $
\phi_2,
\phi_3, \chi$ and $\ox_i$, which we write as,
\bea
P_1&=& \left(\begin{array}{ccc}
1&0&0\\
0&e^{i \chi}&0\\
0&0&1\\
\end{array}\right) \label{P1}\\
P_2&=& \left(\begin{array}{ccc}
1&0&0\\
0&e^{i \phi_2}&0\\
0&0&e^{i \phi_3}\\
\end{array}\right)\\
P_3&=& \left(\begin{array}{ccc}
e^{i \ox_1}&0&0\\
0&e^{i \ox_2}&0\\
0&0&e^{i \ox_3}\\
\end{array}\right)\\ \nonumber
\eea

We briefly summarise the different steps of diagonalisation following \cite{King:2002nf}. We
start by
multiplying the mass matrix, given by Eq.~(\ref{nm}), by the inner complex phase matrices $
P_2$, which helps in performing the diagonalisation procedure. This
process modifies the phases of the matrix as follows,
\be
m_{LL} = \left( \begin{array}{ccc}
|m_{11}|e^{i\phi_{11}}
& |m_{12}|e^{i(\phi_{12}-\phi_2)}
& |m_{13}|e^{i(\phi_{13}-\phi_3)}  \\
|m_{12}|e^{i(\phi_{12}-\phi_2)}
& |m_{22}|e^{i(\phi_{22}-2\phi_2)}
& |m_{23}|e^{i(\phi_{23}-\phi_2-\phi_3)} \\
|m_{13}|e^{i(\phi_{13}-\phi_3)}
& |m_{23}|e^{i(\phi_{23}-\phi_2-\phi_3)}
& |m_{33}|e^{i(\phi_{33}-2\phi_3)}
\end{array}
\right)
\label{nmm}
\ee
After re-phasing the matrix, we proceed by applying the real rotation $R_{23}$, defined in
Eq.~(\ref{R23}). This step modifies the lower 23 block of the mass
matrix by putting zeroes in the 23, 32
elements of the matrix \cite{King:2002nf},
\be \label{diago23}
\left( \begin{array}{cc}
\tilde{m}_{22} &  0\\
0 &  m_3'
\end{array}
\right)
\equiv
{R^{\nu_L}_{23}}^T
\left( \begin{array}{cc}
|m_{22}|e^{i (\phi_{22} -2\phi_2)}
& |m_{23}|e^{i(\phi_{23}-\phi_2-\phi_3)}  \\
|m_{23}|e^{i(\phi_{23}-\phi_2-\phi_3)}
&|m_{33}|e^{i (\phi_{33}- 2\phi_3)}
\end{array}
\right)
R^{\nu_L}_{23}
\ee
This diagonalisation not only modifies the masses $m_{22}$ and $m_{33}$ but also all
the other mass entries except $m_{11}$. Substituting for these masses in terms of the Yukawa couplings, we get an analytic formula for $\tan(\theta_{23})$. The phases $\phi_2, \phi_3$, appearing in Eqs.(\ref{nmm},\ref{diago23}), are fixed in order to have real $\tan(\theta_{23})$. The next step, as shown in Eq.~(\ref{diagon}), is to
apply the rotation $R_{13}$ which diagonalises the outer 13 block. Similar to the previous
step,
this rotation modifies the matrix by putting zeros in the 13, 31 entries.

After applying the 13 rotation, the neutrino mass matrix can be written as,
\be \label{Rot13}
{R_{13}^{\nu_L}}^T{R_{23}^{\nu_L}}^T {P_2^{\nu_L}}^* m_{LL} P_2^{\nu_L}
R_{23}^{\nu_L}
R_{13}^{\nu_L} = \left(\begin{array}{ccc}
\tilde{m}_{11}&\tilde{m}_{12}&0\\
\tilde{m}_{12}&\tilde{m}_{22}&0\\
0&0&m'_3
 \end{array}\right)
\ee
The last step of the diagonalisation involves modifying the upper 12 block of the matrix. To
do this,
we first multiply the result of the last step by the phase matrix ${P_1}$, which introduces the phase $
\chi$.
We then apply the real rotation $R_{12}$. The neutrino mass matrix can then be written as
follows,
\be \label{R12}
{R^\nu_{12}}^T
 \left(\begin{array}{ccc}
\tilde{m}_{11}&\tilde{m}'_{12}&0\\
\tilde{m}'_{12}&\tilde{m}'_{22}&0\\
0&0&m'_3
 \end{array}\right)R^\nu_{12}=  \left(\begin{array}{ccc}
m'_1&0&0\\
0&m'_2&0\\
0&0&m'_3
 \end{array}\right),
\ee
where the new masses $\tilde{m}'_{12}, \tilde{m}'_{22}$ are written as,
\[ \tilde{m}'_{12} = \tilde{m}_{12} ~e^{i \chi},~~ \tilde{m}'_{22} = \tilde{m}_{22}~ e^{i 2 \chi}.\]
From Eq.~(\ref{R12}), we can see that the neutrino matrix is successfully diagonalised,
however, we
still need to multiply the result by the phase matrix $P_3$ in order to make all the
diagonal elements real. To proceed, we write the resulting mass matrix by substituting for
the
diagonal mass terms as $m'_i \equiv m_i e^{i \phi'_i},~i=\{1,2,3\}$. We then apply the phase
matrix
and write the phases $\ox_i$ as $\ox_i = \phi'_i/ 2$. These phases cancel with the phases
of the
neutrino mass matrix which gives a real, diagonal, neutrino matrix as required.

%%%%%%%%%%%%%%%%%%%%%%%%%%%%%%%%%%%%%%%%%%%%
%%%%
\section{Derivation of mass terms} \label{Amass}

In this Appendix, we present the derivations of the mass terms resulting from the
diagonalisation of
the mass matrix. After applying the rotation $R_{23}$ (c.f.\ Eq.~(\ref{R23})), we can derive
expressions for the
masses $m'_3$ and $\tilde{m}_{22}$ which are necessary for deriving expressions for $
\theta_{13}$ and $\tan(\theta_{12})$. To find these masses, we first take the trace of both
sides
of Eq.~(\ref{diago23}) which gives,
\begin{eqnarray} \label{trace}
\tilde{m}_{22} + m'_3 &\approx& m_{22} e^{-i 2 \phi_2}+ m_{33} e^{-i 2 \phi_3}
\nonumber\\
&\approx&e^{ i (2 \phi_e - 2 \phi_2)}\frac{|e|^2 +|f|^2}{Y} \left(1+ \epsilon_3 (s^0_{23})^2 +
(c^0_{23})^2 (\epsilon_2 + \eta_1) \right),
\end{eqnarray}
We can also express the determinant of both sides of Eq.~(\ref{diago23}). This reads,
\bea \label{deter}
\tilde{m}_{22} m'_3 &=& m_{22} e^{-2 i \phi_2} m_{33} e^{-2 i \phi_3} - (m_{23})^2
e^{-2 i (\phi_2+\phi_3)} \nonumber \\
&= &e^{-2 i (\phi_2+\phi_3)} \left( \frac{c'^2}{X'} (\frac{b^2}{X}+ \frac{e^2}{Y}) + \frac{(b f - e
c)^2}{X
Y} \right)
\eea \label{m22n}

We take the mass term $m'_3 $ to have the following form,
\be \label{m3pn}
m'_3 \approx m'^0_3 (1+\beta),
\ee
where the parameter $\beta$ is given by Eq.~(\ref{beta}) and the mass term $m'^0_3 \equiv
m'_3|
_{\epsilon_i=0,\eta_i =0}$ is given by,
\be \label{m3pz}
m'^0_3 \approx e^{ i (2 \phi_e - 2 \phi_2)}\frac{|e|^2 +|f|^2}{Y},
\ee
Using Eqs.(\ref{trace}), (\ref{deter}), (\ref{m3pz}), $\tilde{m}_{22}$ can be written as,
\be \label{m22c}
\tilde{m}_{22} \approx \left(\tilde{m}_{22}^0 + \eta_2 \left(\frac{b^2}{X}+\frac{e^2}{Y}\right)
\right)  (1-
\beta),
\ee
where
\be
\tilde{m}_{22}^0 \equiv \tilde{m}_{22}|_{c' =0,\epsilon_i = 0} \approx e^{-2 i \phi_2} \frac{(b
c_{23}^0- c s_{23}^0 e^{i (\phi_e- \phi_f)})^2}{X},
\ee
and the parameter $\eta_2$ is given by Eq.~(\ref{eta2}).

In addition to the derivation of the masses $\tilde{m}_{22}, m'_3$, applying the rotation
$R_{23}$,
modifies the masses $m_{12}, m_{13}$. These become $\tilde{m}_{12}, \tilde{m}_{13}$,
after
diagonalising the 23 block, and can be derived to leading order as follows
\bea \label{m12c}
\tilde{m}_{12} &=& c_{23} m_{12} e^{-i \phi_2} - s_{23} m_{13} e^{-i \phi_3}, \nonumber \\
&\approx& \tilde{ m}_{12}^0 (1 - \gamma (s_{23}^0)^2)-e^{-i \phi_3} \gamma s_{23}^0
(\frac{a c}{X}
+ \frac{d f}{Y}),
\eea
\bea  \label{m13c}
\tilde{m}_{13} &=& s_{23} m_{12} e^{-i \phi_2} + c_{23} m_{13} e^{-i \phi_3} \nonumber \\
&\approx& \tilde{m}_{13}^0 (1 - \gamma (s_{23}^0)^2) + e^{-i \phi_2} \gamma
s_{23}^0(\frac{a b}{X}
+ \frac{d e}{Y}),
\eea
where the parameter $\gamma$ is given by Eq.~(\ref{gamma}), the masses $
\tilde{m}_{12}^0 \equiv
\tilde{m}_{12}|_{c'=0,\epsilon_i=0} $ and $\tilde{m}_{13}^0 \equiv \tilde{m}_{13}|
_{c'=0,\epsilon_i=0} $ are given to leading order, as presented in \cite{King:2002nf}, by
\bea
\tilde{m}_{12}^0 &\approx& e^{-i \phi_2}\frac{a (c_{23}^0 b -s_{23}^0 c e^{i (\phi_e -
\phi_f)})}{X},\\
\tilde{m}_{13}^0 &\approx& e^{-i \phi_2}\left( \frac{a}{X} (s^0_{23} b + c^0_{23} c
e^{i(\phi_e-
\phi_f)})+ e^{i \phi_e} \frac{d \sqrt{|e|^2+|f|^2}}{Y} \label{m13z}\right).
\eea

After applying the $R_{13}$ rotation, we obtain another mass term, $\tilde{m}_{11}$, which
can be
presented to leading order as
\bea \label{m11c}
\tilde{m}_{11} &\approx& m_{11}- \frac{\tilde{m}_{13}^2}{m'_3} \nonumber \\
&\approx& \tilde{m}^0_{11} (1 - 2 \gamma (s^0_{23})^2 - \beta)+\frac{a^2}{X}(2 \gamma
(s^0_{23})^2
+ \beta)+\frac{d^2}{Y} ( 2 \gamma (s^0_{23})^2 + \beta)  \nonumber\\
&-&2 e^{-2 i \phi_e} \epsilon_6 (s^0_{23} b + c^0_{23} c e^{i (\phi_e - \phi_f)}) \gamma
s^0_{23}
\left( \frac{a b}{X}+\frac{d e}{Y} \right)
\eea
where the leading order form of  $\tilde{m}^0_{11} \equiv \tilde{m}_{11}|_{\eta_i =0,
\epsilon_i =0}$
is given in
 \cite{King:2002nf},
\be
{m}^0_{11} \approx \frac{a^2}{X} - e^{-i \phi_e} \frac{2 d}{ \sqrt{|e|^2+ |f|^2}}\frac{a (s^0_{23}
b +
c^0_{23} c e^{i (\phi_e - \phi_f)})}{X},
\ee
The small parameter $\epsilon_6$ is written as,
\[\epsilon_6 = \frac{a}{X}\left( \frac{|e|^2+|f|^2}{Y}\right)^{-1}.\]

Similar to the derivation of the masses $\tilde{m}_{22}, m'_3$, the neutrino masses $m'_1$
and
$m'_2$ can be
written using the trace and the determinant of the upper 12 block of Eq.~(\ref{R12}). The
real neutrino
masses $m_1, m_2$ can then be written, in the SD cases, as
\bea
m_1 &\approx&  \frac{|c'|^2}{6 X'} \left(1- \frac{Y}{X}\frac{|b|^2}{|e|^2} + \frac{2 |d|}{|e|}\right),
\label{m1}\\
m_2 &\approx&\left( \frac{3 |b|^2}{X} +\frac{|c'|^2}{3 X'} \left(1- \frac{Y}{X}\frac{|b|^2}{|e|^2}-
\frac{|d|}{|
e|}\right) \cos(2 \phi_{c'}) \right),\label{m2}
\eea

The neutrino mass $ m_3$ can be written in the SD cases, using Eqs.(\ref{m3pn}),
(\ref{m3pz}), as
\be \label{m3}
m_3 \approx \left(\frac{2 |e|^2 }{Y} + \frac{|c'|^2}{X'} \cos (2 \phi_{c'}) \right).
\ee


\begin{thebibliography}{10}


\bibitem{seesaw}
P.~Minkowski,
  %``Mu $\to$ E Gamma At A Rate Of One Out Of 1-Billion Muon Decays?,''
  Phys.\ Lett.\ B {\bf 67} (1977) 421;
  %%CITATION = PHLTA,B67,421;%%
M. Gell-Mann, P. Ramond and R. Slansky in Sanibel Talk,
CALT-68-709, Feb 1979, and in {\it Supergravity} (North Holland,
Amsterdam 1979);
T. Yanagida in {\it Proc. of the Workshop on Unified Theory and
Baryon Number of the Universe}, KEK, Japan, 1979;
S.L.Glashow, Cargese Lectures (1979);
%\cite{Mohapatra:1979ia}
%\bibitem{Mohapatra:1979ia}
R.~N.~Mohapatra and G.~Senjanovic,
%``Neutrino Mass And Spontaneous Parity Nonconservation,''
Phys.\ Rev.\ Lett.\  {\bf 44} (1980) 912;
%%CITATION = PRLTA,44,912;%%
J.~Schechter and J.~W.~Valle,
%``Neutrino Decay And Spontaneous Violation Of Lepton Number,''
Phys.\ Rev.\ D {\bf 25} (1982) 774.
%%CITATION = PHRVA,D25,774;%%

%\cite{King:1998jw}
\bibitem{King:1998jw}
S.~F.~King,
%``Atmospheric and solar neutrinos with a heavy singlet,''
Phys.\ Lett.\ B {\bf 439} (1998) 350
[arXiv:hep-ph/9806440];
%%CITATION = HEP-PH 9806440;%%
%\cite{King:1999cm}
%\bibitem{King:1999cm}
  S.~F.~King,
  %``Atmospheric and solar neutrinos from single right-handed neutrino
  %dominance and U(1) family symmetry,''
  Nucl.\ Phys.\  B {\bf 562} (1999) 57
  [arXiv:hep-ph/9904210];
  %%CITATION = NUPHA,B562,57;%%
%\cite{Altarelli:1999dg}
%\bibitem{Altarelli:1999dg}
  G.~Altarelli, F.~Feruglio and I.~Masina,
  %``Large neutrino mixing from small quark and lepton mixings,''
  Phys.\ Lett.\  B {\bf 472} (2000) 382
  [arXiv:hep-ph/9907532].
  %%CITATION = PHLTA,B472,382;%%



\bibitem{King:1999mb}
  S.~F.~King,
  %``Large mixing angle MSW and atmospheric neutrinos from single  right-handed
  %neutrino dominance and U(1) family symmetry,''
  Nucl.\ Phys.\  B {\bf 576} (2000) 85
  [arXiv:hep-ph/9912492].
  %%CITATION = NUPHA,B576,85;%%
  
%\cite{Lavignac:2002gf}
\bibitem{Lavignac:2002gf}
  S.~Lavignac, I.~Masina and C.~A.~Savoy,
  %``Large solar angle and seesaw mechanism: A bottom-up perspective,''
  Nucl.\ Phys.\  B {\bf 633} (2002) 139
  [arXiv:hep-ph/0202086].
  %%CITATION = NUPHA,B633,139;%%

%\cite{King:2002nf}
\bibitem{King:2002nf}
  S.~F.~King,
  %``Constructing the large mixing angle MNS matrix in see-saw models with
  %right-handed neutrino dominance,''
  JHEP {\bf 0209} (2002) 011
  [arXiv:hep-ph/0204360].
  %%CITATION = JHEPA,0209,011;%%



  %\cite{King:2002qh}
\bibitem{King:2002qh}
  S.~F.~King,
  %``Leptogenesis - MNS link in unified models with natural neutrino mass
  %hierarchy,''
  Phys.\ Rev.\  D {\bf 67} (2003) 113010
  [arXiv:hep-ph/0211228].
  %%CITATION = PHRVA,D67,113010;%%

\bibitem{King:2004}
 For a review of sequential dominance, see: S.~Antusch and S.~F.~King,
  %``Sequential dominance,''
  New J.\ Phys.\  {\bf 6} (2004) 110
  [arXiv:hep-ph/0405272];
  %%CITATION = HEP-PH 0405272;%%

%\cite{Haba:2008dp}
\bibitem{Haba:2008dp}
  N.~Haba, R.~Takahashi, M.~Tanimoto and K.~Yoshioka,
  %``Tri-bimaximal Mixing from Cascades,''
  Phys.\ Rev.\  D {\bf 78} (2008) 113002
  [arXiv:0804.4055 [hep-ph]].
  %%CITATION = PHRVA,D78,113002;%%



%\cite{Bandyopadhyay:2007kx}
\bibitem{Bandyopadhyay:2007kx}
  A.~Bandyopadhyay {\it et al.}  [ISS Physics Working Group],
  %``Physics at a future Neutrino Factory and super-beam facility,''
  Rept.\ Prog.\ Phys.\  {\bf 72} (2009) 106201
  [arXiv:0710.4947 [hep-ph]].
  %%CITATION = RPPHA,72,106201;%%


 %\cite{Chen:2009um}
\bibitem{Chen:2009um}
  M.~C.~Chen and S.~F.~King,
  %``$A_4$ See-Saw Models and Form Dominance,''
  arXiv:0903.0125 [hep-ph].
  %%CITATION = ARXIV:0903.0125;%%

%\cite{Antusch:2007jd}
\bibitem{Antusch:2007jd}
  S.~Antusch, L.~E.~Ibanez and T.~Macri,
  %``Neutrino Masses and Mixings from String Theory Instantons,''
  JHEP {\bf 0709} (2007) 087
  [arXiv:0706.2132 [hep-ph]].
  %%CITATION = JHEPA,0709,087;%%

%\cite{King:2005bj}
\bibitem{King:2005bj}
  S.~F.~King,
  %``Predicting neutrino parameters from SO(3) family symmetry and  quark-lepton
  %unification,''
  JHEP {\bf 0508} (2005) 105
  [arXiv:hep-ph/0506297].
  %%CITATION = JHEPA,0508,105;%%


\bibitem{tribi}
%\cite{Harrison:2002er}
%bibitem{Harrison:2002er}
P.~F.~Harrison, D.~H.~Perkins and W.~G.~Scott,
%``Tri-bimaximal mixing and the neutrino oscillation data,''
Phys.\ Lett.\ B {\bf 530} (2002) 167
[arXiv:hep-ph/0202074];
%%CITATION = HEP-PH 0202074;%%
%\cite{Harrison:2002kp}
%\bibitem{Harrison:2002kp}
P.~F.~Harrison and W.~G.~Scott,
%``Symmetries and generalisations of tri-bimaximal neutrino mixing,''
Phys.\ Lett.\ B {\bf 535} (2002) 163
[arXiv:hep-ph/0203209];
%%CITATION = HEP-PH 0203209;%%
%\cite{Harrison:2003aw}
%\bibitem{Harrison:2003aw}
P.~F.~Harrison and W.~G.~Scott,
%``Permutation symmetry, tri-bimaximal neutrino mixing and the S3 group
%characters,''
Phys.\ Lett.\ B {\bf 557} (2003) 76
[arXiv:hep-ph/0302025];
see also %\cite{Wolfenstein:1978uw}
%bibitem{Wolfenstein:1978uw}
L.~Wolfenstein,
%``Oscillations Among Three Neutrino Types And CP Violation,''
Phys.\ Rev.\ D {\bf 18} (1978) 958.
%%CITATION = PHRVA,D18,958;%%




%%%%%%%%%%%%%%%%%%%%%%%%%%%%%%%%%%%%%%%%%%%%%%%%%%%%%%%%%%%%%%%%%%%%%%%%%%%%%

%S4 is minimal for TB mixing
\bibitem{Lam}
C.~S.~Lam,
 Phys.\ Lett.\  B {\bf 656} (2007) 193 [arXiv:0708.3665];~\\
C.~S.~Lam,
  Phys.\ Rev.\ Lett.\  {\bf 101} (2008) 121602 [arXiv:0804.2622];~\\
C.~S.~Lam,
  Phys.\ Rev.\  D {\bf 78} (2008) 073015 [arXiv:0809.1185].



%%%%%%%%%%%%%%%%%%%%%%%%%%%%%%%%%%%%%%%%%%%%%%%%%%%%%%%%%%%%%%%%%%%%%%%%%%%%%

   %\cite{King:2009ap}
\bibitem{King:2009ap}
  S.~F.~King and C.~Luhn,
  %``On the origin of neutrino flavour symmetry,''
  JHEP {\bf 0910} (2009) 093
  [arXiv:0908.1897 [hep-ph]].
  %%CITATION = JHEPA,0910,093;%%

%%%%%%%%%%%%%%%%%%%%%%%%%%%%%%%%%%%%%%%

%S3
\bibitem{S3-L}
W.~Grimus and L.~Lavoura,
  JHEP {\bf 0508} (2005) 013 [hep-ph/0504153];~\\
W.~Grimus and L.~Lavoura,
  JHEP {\bf 0601} (2006) 018  [hep-ph/0509239];~\\
R.~N.~Mohapatra, S.~Nasri and H.~B.~Yu,
  Phys.\ Lett.\  B {\bf 639} (2006) 318 [hep-ph/0605020];~\\
Y.~Koide,
  Eur.\ Phys.\ J.\  C {\bf 50} (2007) 809 [hep-ph/0612058];~\\
M.~Mitra and S.~Choubey,
  Phys.\ Rev.\  D {\bf 78} (2008) 115014 [arXiv:0806.3254].

%%%%%%%%%%%%%%%%%%%%%%%%%%%%%%%%%%%%%%%%%%%%%%%%%%%%%%%%%%%%%%%%%%%%%%%%%%%%%

%D4
\bibitem{Dn-L}
 W.~Grimus and L.~Lavoura,
  Phys.\ Lett.\  B {\bf 572} (2003) 189
  [hep-ph/0305046];~\\
A.~Adulpravitchai, A.~Blum and C.~Hagedorn,
  JHEP {\bf 0903} (2009) 046  [arXiv:0812.3799];
%\cite{Ishimori:2008gp}
%\bibitem{Ishimori:2008gp}
  H.~Ishimori, T.~Kobayashi, H.~Ohki, Y.~Omura, R.~Takahashi and M.~Tanimoto,
  %``D4 Flavor Symmetry for Neutrino Masses and Mixing,''
  Phys.\ Lett.\  B {\bf 662} (2008) 178
  [arXiv:0802.2310 [hep-ph]];
  %%CITATION = PHLTA,B662,178;%%
%\cite{Ishimori:2008ns}
%\bibitem{Ishimori:2008ns}
  H.~Ishimori, T.~Kobayashi, H.~Ohki, Y.~Omura, R.~Takahashi and M.~Tanimoto,
  %``Soft supersymmetry breaking terms from D4 x Z2 lepton flavor symmetry,''
  Phys.\ Rev.\  D {\bf 77} (2008) 115005
  [arXiv:0803.0796 [hep-ph]].
  %%CITATION = PHRVA,D77,115005;%%


  
  
  

%%%%%%%%%%%%%%%%%%%%%%%%%%%%%%%%%%%%%%%%%%%%%%%%%%%%%%%%%%%%%%%%%%%%%%%%%%%%%

%A4
\bibitem{A4-L}
E.~Ma and G.~Rajasekaran,
   Phys.\ Rev.\  D {\bf 64} (2001) 113012 [hep-ph/0106291];~\\
E.~Ma,
  Phys.\ Rev.\  D {\bf 73} (2006) 057304 [hep-ph/0511133];~\\
G.~Altarelli and F.~Feruglio,
  Nucl.\ Phys.\  B {\bf 720} (2005) 64
  [hep-ph/0504165];~\\
K.~S.~Babu and X.~G.~He,
  hep-ph/0507217;~\\
G.~Altarelli and F.~Feruglio,
  Nucl.\ Phys.\  B {\bf 741} (2006) 215
  [hep-ph/0512103];~\\
G.~Altarelli, F.~Feruglio and Y.~Lin,
  Nucl.\ Phys.\  B {\bf 775} (2007) 31
  [hep-ph/0610165];~\\
M.~Hirsch, A.~S.~Joshipura, S.~Kaneko and J.~W.~F.~Valle,
  Phys.\ Rev.\ Lett.\  {\bf 99} (2007) 151802 [hep-ph/0703046];~\\
M.~Honda and M.~Tanimoto,
  Prog.\ Theor.\ Phys.\  {\bf 119} (2008) 583 [arXiv:0801.0181];~\\
Y.~Lin,
  Nucl.\ Phys.\  B {\bf 813} (2009) 91  [arXiv:0804.2867];~\\
M.~C.~Chen and S.~F.~King,
  JHEP {\bf 0906} (2009) 072
  [arXiv:0903.0125];~\\
G.~Altarelli and D.~Meloni,
  J.\ Phys.\ G {\bf 36} (2009) 085005
  [arXiv:0905.0620].

%%%%%%%%%%%%%%%%%%%%%%%%%%%%%%%%%%%%%%%%%%%%%%%%%%%%%%%%%%%%%%%%%%%%%%%%%%%%%

%S4
\bibitem{S4-L}
Y.~Koide,
  JHEP {\bf 0708} (2007) 086 [arXiv:0705.2275].

%%%%%%%%%%%%%%%%%%%%%%%%%%%%%%%%%%%%%%%%%%%%%%%%%%%%%%%%%%%%%%%%%%%%%%%%%%%%%

%delta54
\bibitem{delta54-L}
H.~Ishimori, T.~Kobayashi, H.~Okada, Y.~Shimizu and M.~Tanimoto,
  JHEP {\bf 0904} (2009) 011  [arXiv:0811.4683].

%%%%%%%%%%%%%%%%%%%%%%%%%%%%%%%%%%%%%%%%%%%%%%%%%%%%%%%%%%%%%%%%%%%%%%%%%%%%%

%S3-LQ
\bibitem{S3-LQ}
J.~Kubo, A.~Mondragon, M.~Mondragon and E.~Rodriguez-Jauregui,
   Prog.\ Theor.\ Phys.\  {\bf 109} (2003) 795,
   Erratum-ibid.\  {\bf 114} (2005) 287 [hep-ph/0302196];~\\
S.~L.~Chen, M.~Frigerio and E.~Ma,
   Phys.\ Rev.\  D {\bf 70} (2004) 073008
  [Erratum-ibid.\  D {\bf 70} (2004) 079905]
  [hep-ph/0404084];~\\
F.~Feruglio and Y.~Lin,
  Nucl.\ Phys.\  B {\bf 800} (2008) 77  [arXiv:0712.1528].

%%%%%%%%%%%%%%%%%%%%%%%%%%%%%%%%%%%%%%%%%%%%%%%%%%%%%%%%%%%%%%%%%%%%%%%%%%%%%

%Dn-Q
\bibitem{Dn-LQ}
A.~Blum, C.~Hagedorn and M.~Lindner,
  Phys.\ Rev.\  D {\bf 77} (2008) 076004 [arXiv:0709.3450];~\\
A.~Blum, C.~Hagedorn and A.~Hohenegger,
  JHEP {\bf 0803} (2008) 070  [arXiv:0710.5061];~\\
A.~Blum and C.~Hagedorn,
  Nucl.\ Phys.\  B {\bf 821} (2009) 327
  [arXiv:0902.4885].

%%%%%%%%%%%%%%%%%%%%%%%%%%%%%%%%%%%%%%%%%%%%%%%%%%%%%%%%%%%%%%%%%%%%%%%%%%%%%

%Q6, Dicyclic
\bibitem{Q6-LQ}
M.~Frigerio, S.~Kaneko, E.~Ma and M.~Tanimoto,
  Phys.\ Rev.\  D {\bf 71} (2005) 011901  [hep-ph/0409187];~\\
K.~S.~Babu and J.~Kubo,
  Phys.\ Rev.\  D {\bf 71} (2005) 056006 [hep-ph/0411226];~\\
Y.~Kajiyama, E.~Itou and J.~Kubo,
  Nucl.\ Phys.\  B {\bf 743} (2006) 74 [hep-ph/0511268].

%%%%%%%%%%%%%%%%%%%%%%%%%%%%%%%%%%%%%%%%%%%%%%%%%%%%%%%%%%%%%%%%%%%%%%%%%%%%%

%A4-Q
\bibitem{A4-LQ}
K.~S.~Babu, E.~Ma and J.~W.~F.~Valle,
  Phys.\ Lett.\  B {\bf 552} (2003) 207  [hep-ph/0206292];~\\
F.~Bazzocchi, S.~Morisi and M.~Picariello,
  Phys.\ Lett.\  B {\bf 659} (2008) 628 [arXiv:0710.2928].

%%%%%%%%%%%%%%%%%%%%%%%%%%%%%%%%%%%%%%%%%%%%%%%%%%%%%%%%%%%%%%%%%%%%%%%%%%%%%

%A4-PS
\bibitem{A4-LQ:King}
S.~F.~King and M.~Malinsky,
  Phys.\ Lett.\  B {\bf 645} (2007) 351 [hep-ph/0610250].

%%%%%%%%%%%%%%%%%%%%%%%%%%%%%%%%%%%%%%%%%%%%%%%%%%%%%%%%%%%%%%%%%%%%%%%%%%%%%

%T'-Q
\bibitem{doubleA4-LQ}
A.~Aranda, C.~D.~Carone and R.~F.~Lebed,
  Phys.\ Rev.\  D {\bf 62} (2000) 016009  [hep-ph/0002044];~\\
F.~Feruglio, C.~Hagedorn, Y.~Lin and L.~Merlo,
  Nucl.\ Phys.\  B {\bf 775} (2007) 120   [hep-ph/0702194];~\\
M.~C.~Chen and K.~T.~Mahanthappa,
  Phys.\ Lett.\  B {\bf 652} (2007) 34 [arXiv:0705.0714];~\\
P.~H.~Frampton and T.~W.~Kephart,
  JHEP {\bf 0709} (2007) 110  [arXiv:0706.1186];~\\
A.~Aranda,
  Phys.\ Rev.\  D {\bf 76} (2007) 111301  [arXiv:0707.3661];~\\
P.~H.~Frampton and S.~Matsuzaki,
  Phys.\ Lett.\  B {\bf 679} (2009) 347
  [arXiv:0902.1140].

%%%%%%%%%%%%%%%%%%%%%%%%%%%%%%%%%%%%%%%%%%%%%%%%%%%%%%%%%%%%%%%%%%%%%%%%%%%%%

%S4xSM
\bibitem{S4-LQ:Hagedorn}
C.~Hagedorn, M.~Lindner and R.~N.~Mohapatra,
  JHEP {\bf 0606} (2006) 042 [hep-ph/0602244];~\\
G.~J.~Ding,
  Nucl.\ Phys.\  B {\bf 827} (2010) 82
  [arXiv:0909.2210];~\\
D.~Meloni,
  %``A See-Saw $S_4$ model for fermion masses and mixings,''
  arXiv:0911.3591 [hep-ph].


%%%%%%%%%%%%%%%%%%%%%%%%%%%%%%%%%%%%%%%%%%%%%%%%%%%%%%%%%%%%%%%%%%%%%%%%%%%%%

%S4xSM, but our group theory
\bibitem{S4-group}
F.~Bazzocchi, L.~Merlo and S.~Morisi,
  %``Fermion Masses and Mixings in a S4-based Model,''
  Nucl.\ Phys.\  B {\bf 816} (2009) 204
  [arXiv:0901.2086].

%%%%%%%%%%%%%%%%%%%%%%%%%%%%%%%%%%%%%%%%%%%%%%%%%%%%%%%%%%%%%%%%%%%%%%%%%%%%%

%A4-SU(5)
\bibitem{A4-SU5}
G.~Altarelli, F.~Feruglio and C.~Hagedorn,
  JHEP {\bf 0803} (2008) 052
  [arXiv:0802.0090];~\\
P.~Ciafaloni, M.~Picariello, E.~Torrente-Lujan and A.~Urbano,
  Phys.\ Rev.\  D {\bf 79} (2009) 116010
  [arXiv:0901.2236];~\\
T.~J.~Burrows and S.~F.~King,
  arXiv:0909.1433.

%%%%%%%%%%%%%%%%%%%%%%%%%%%%%%%%%%%%%%%%%%%%%%%%%%%%%%%%%%%%%%%%%%%%%%%%%%%%%

%A4-SO(10)
\bibitem{A4-LQ:Morisi}
S.~Morisi, M.~Picariello and E.~Torrente-Lujan,
  Phys.\ Rev.\  D {\bf 75} (2007) 075015  [hep-ph/0702034];~\\
F.~Bazzocchi, M.~Frigerio and S.~Morisi,
  Phys.\ Rev.\  D {\bf 78} (2008) 116018  [arXiv:0809.3573].

%%%%%%%%%%%%%%%%%%%%%%%%%%%%%%%%%%%%%%%%%%%%%%%%%%%%%%%%%%%%%%%%%%%%%%%%%%%%%

%S4 x SU(5) ... another try
\bibitem{S4-LQ}
H.~Ishimori, Y.~Shimizu and M.~Tanimoto,
  %``S4 Flavor Symmetry of Quarks and Leptons in SU(5) GUT,''
  Prog.\ Theor.\ Phys.\  {\bf 121} (2009) 769
  [arXiv:0812.5031];~\\
H.~Ishimori, T.~Kobayashi, H.~Ohki, H.~Okada, Y.~Shimizu and M.~Tanimoto,
  %``Non-Abelian Discrete Symmetries in Particle Physics,''
  arXiv:1003.3552;~\\
%\cite{Hagedorn:2010th}
%\bibitem{Hagedorn:2010th}
  C.~Hagedorn, S.~F.~King and C.~Luhn,
  %``A SUSY GUT of Flavour with S4 x SU(5) to NLO,''
  arXiv:1003.4249 [Unknown].
  %%CITATION = ARXIV:1003.4249;%%





%%%%%%%%%%%%%%%%%%%%%%%%%%%%%%%%%%%%%%%%%%%%%%%%%%%%%%%%%%%%%%%%%%%%%%%%%%%%%

%S4-PS/SO(10)
\bibitem{Mohapatra:2003tw}
  D.~G.~Lee and R.~N.~Mohapatra,
  Phys.\ Lett.\  B {\bf 329} (1994) 463
  [hep-ph/9403201];~\\
  R.~N.~Mohapatra, M.~K.~Parida and G.~Rajasekaran,
  Phys.\ Rev.\  D {\bf 69} (2004) 053007
  [hep-ph/0301234];~\\
  Y.~Cai and H.~B.~Yu,
  Phys.\ Rev.\  D {\bf 74} (2006) 115005
  [hep-ph/0608022];~\\
  M.~K.~Parida,
  Phys.\ Rev.\  D {\bf 78} (2008) 053004
  [arXiv:0804.4571];~\\
  B.~Dutta, Y.~Mimura and R.~N.~Mohapatra,
  arXiv:0911.2242.

%%%%%%%%%%%%%%%%%%%%%%%%%%%%%%%%%%%%%%%%%%%%%%%%%%%%%%%%%%%%%%%%%%%%%%%%%%%%%

%PSL(2,7) xSO(10)
\bibitem{King:2009mk}
S.~F.~King and C.~Luhn,
  %``A new family symmetry for SO(10) GUTs,''
  Nucl.\ Phys.\  B {\bf 820} (2009) 269
  [arXiv:0905.1686];~\\
S.~F.~King and C.~Luhn,
  %``A Supersymmetric Grand Unified Theory of Flavour with PSL(2,7) x SO(10),''
  Nucl.\ Phys.\  B {\bf 832} (2010) 414
 [arXiv:0912.1344].


%%%%%%%%%%%%%%%%%%%%%%%%%%%%%%%%%%%%%%%%%%%%%%%%%%%%%%%%%%%%%%%%%%%%%%%%%%%%%

% Z7 x Z3 SO(10) (also Sigma(81))
\bibitem{Z7Z3-LQ}
C.~Luhn, S.~Nasri and P.~Ramond,
  Phys.\ Lett.\  B {\bf 652} (2007) 27 [arXiv:0706.2341].

\bibitem{T7-LQ}
C.~Hagedorn, M.~A.~Schmidt and A.~Y.~Smirnov,
  Phys.\ Rev.\  D {\bf 79} (2009) 036002 [arXiv:0811.2955].


%%%%%%%%%%%%%%%%%%%%%%%%%%%%%%%%%%%%%%%%%%%%%%%%%%%%%%%%%%%%%%%%%%%%%%%%%%%%%

%Delta(27) SO(10)/PS
\bibitem{delta27-LQ:King}
I.~de Medeiros Varzielas, S.~F.~King and G.~G.~Ross,
   Phys.\ Lett.\  B {\bf 644} (2007) 153
  [hep-ph/0512313];~\\
I.~de Medeiros Varzielas, S.~F.~King and G.~G.~Ross,
  Phys.\ Lett.\  B {\bf 648} (2007) 201  [hep-ph/0607045];~\\
F.~Bazzocchi and I.~de Medeiros Varzielas,
  Phys.\ Rev.\  D {\bf 79} (2009) 093001
  [arXiv:0902.3250].

%%%%%%%%%%%%%%%%%%%%%%%%%%%%%%%%%%%%%%%%%%%%%%%%%%%%%%%%%%%%%%%%%%%%%%%%%%%%%

%SO(3)
\bibitem{SO(3)-LQ:King}
S.~F.~King and M.~Malinsky,
  JHEP {\bf 0611} (2006) 071 [hep-ph/0608021].

%%%%%%%%%%%%%%%%%%%%%%%%%%%%%%%%%%%%%%%%%%%%%%%%%%%%%%%%%%%%%%%%%%%%%%%%%%%%%

%SU(3)
\bibitem{SU(3)-LQ:Ross}
S.~F.~King and G.~G.~Ross,
  Phys.\ Lett.\ B \textbf{520} (2001) 243 [hep-ph/0108112];~\\
S.~F.~King and G.~G.~Ross,
  Phys.\ Lett.\ B \textbf{574} (2003) 239 [hep-ph/0307190];~\\
I.~de Medeiros Varzielas and G.~G.~Ross,
   Nucl.\ Phys.\  B {\bf 733} (2006) 31 [hep-ph/0507176].

%%%%%%%%%%%%%%%%%%%%%%%%%%%%%%%%%%%%%%%%%%%%%%%%%%%%%%%%%%%%%%%%%%%%%%%%%%%%%

%review of models
\bibitem{Reviews}
S.~F.~King,
  Rept.\ Prog.\ Phys.\ \textbf{67} (2004) 107 [hep-ph/0310204];~\\
R.~N.~Mohapatra {\it et al.},
  Rept.\ Prog.\ Phys.\  {\bf 70} (2007) 1757 [hep-ph/0510213];~\\
R.~N.~Mohapatra and A.~Y.~Smirnov,
  Ann.\ Rev.\ Nucl.\ Part.\ Sci.\  {\bf 56} (2006) 569 [hep-ph/0603118];~\\
C.~H.~Albright,
  arXiv:0905.0146;~\\
G.~Altarelli and F.~Feruglio,
  arXiv:1002.0211.

%%%%%%%%%%%%%%%%%%%%%%%%%%%%%%%%%%%%%%%%%%%%%%%%%%%%%%%%%%%%%%%%%%%%%%%%%%%%%



  %\cite{Fogli:2009ce}
\bibitem{Fogli:2009ce}
  G.~L.~Fogli, E.~Lisi, A.~Marrone, A.~Palazzo and A.~M.~Rotunno,
  %``SNO, KamLAND and neutrino oscillations: theta(13),''
  arXiv:0905.3549 [hep-ph];
  %%CITATION = ARXIV:0905.3549;%%
 %\cite{Fogli:2008cx}
%\bibitem{Fogli:2008cx}
  G.~L.~Fogli, E.~Lisi, A.~Marrone, A.~Palazzo and A.~M.~Rotunno,
  %``What we (would like to) know about the neutrino mass,''
  arXiv:0809.2936 [hep-ph].
  %%CITATION = ARXIV:0809.2936;%%



\bibitem{sumrule}
S.~F.~King,
%``Predicting neutrino parameters from SO(3) family symmetry and quark-lepton
%unification,''
JHEP {\bf 0508} (2005) 105;
%[arXiv:hep-ph/0506297];
I.~Masina,
  %``A maximal atmospheric mixing from a maximal CP violating phase,''
  Phys.\ Lett.\  B {\bf 633} (2006) 134;
%[arXiv:hep-ph/0508031];
  %%CITATION = PHLTA,B633,134;%%
%%CITATION = HEP-PH 0506297;%%
S.~Antusch and S.~F.~King,
%   ``Charged lepton corrections to neutrino mixing angles and CP phases
  %revisited,''
  Phys.\ Lett.\ B {\bf 631} (2005) 42;
%[arXiv:hep-ph/0508044];
  %%CITATION = HEP-PH 0508044;%%
S.~Antusch, P.~Huber, S.~F.~King and T.~Schwetz,
  %``Neutrino mixing sum rules and oscillation experiments,''
  JHEP {\bf 0704} (2007) 060.
%  [arXiv:hep-ph/0702286].
  %%CITATION = JHEPA,0704,060;%%

\bibitem{RG}
%\bibitem{Antusch:2003kp}
  S.~Antusch, J.~Kersten, M.~Lindner and M.~Ratz,
  %``Running neutrino masses, mixings and CP phases: Analytical results and
  %phenomenological consequences,''
  Nucl.\ Phys.\  B {\bf 674} (2003) 401;


%\cite{Dighe:2006sr}
\bibitem{Dighe:2006sr}
  A.~Dighe, S.~Goswami and W.~Rodejohann,
  %``Corrections to Tri-bimaximal Neutrino Mixing: Renormalization and Planck
  %Scale Effects,''
  Phys.\ Rev.\  D {\bf 75} (2007) 073023
  [arXiv:hep-ph/0612328].
  %%CITATION = PHRVA,D75,073023;%%

  %\cite{Boudjemaa:2008jf}
\bibitem{Boudjemaa:2008jf}
  S.~Boudjemaa and S.~F.~King,
  %``Deviations from Tri-bimaximal Mixing: Charged Lepton Corrections and
  %Renormalization Group Running,''
  Phys.\ Rev.\  D {\bf 79} (2009) 033001
  [arXiv:0808.2782 [hep-ph]].
  %%CITATION = PHRVA,D79,033001;%%

%\cite{Antusch:2007ib}
\bibitem{Antusch:2007ib}
  S.~Antusch, S.~F.~King and M.~Malinsky,
  %``Third Family Corrections to Tri-bimaximal Lepton Mixing and a New Sum
  %Rule,''
  Phys.\ Lett.\  B {\bf 671} (2009) 263
  [arXiv:0711.4727 [hep-ph]];
  %%CITATION = PHLTA,B671,263;%%

\bibitem{cnorm}
  S.~Antusch, S.~F.~King and M.~Malinsky,
  %``Third Family Corrections to Quark and Lepton Mixing in SUSY Models with
  %non-Abelian Family Symmetry,''
  JHEP {\bf 0805} (2008) 066
  [arXiv:0712.3759 [hep-ph]].
  %%CITATION = JHEPA,0805,066;%%

 %\cite{King:2009qh}
\bibitem{King:2009qh}
  S.~F.~King,
  %``Tri-bimaximal Neutrino Mixing and $\theta_{13}$,''
  arXiv:0903.3199 [hep-ph].
  %%CITATION = ARXIV:0903.3199;%%

%\cite{King:2007pr}
\bibitem{King:2007pr}
  S.~F.~King,
  %``Parametrizing the lepton mixing matrix in terms of deviations from
  %tri-bimaximal mixing,''
  Phys.\ Lett.\  B {\bf 659} (2008) 244
  [arXiv:0710.0530 [hep-ph]].
  %%CITATION = PHLTA,B659,244;%%

 %\cite{Antusch:2005gp}
\bibitem{Antusch:2005gp}
  S.~Antusch, J.~Kersten, M.~Lindner, M.~Ratz and M.~A.~Schmidt,
  %``Running neutrino mass parameters in see-saw scenarios,''
  JHEP {\bf 0503} (2005) 024
  [arXiv:hep-ph/0501272].
  %%CITATION = JHEPA,0503,024;%%

%\cite{Howl:2009ds}
\bibitem{Howl:2009ds}
  R.~Howl and S.~F.~King,
  %``Solving the Flavour Problem in Supersymmetric Standard Models with Three
  %Higgs Families,''
  arXiv:0908.2067 [hep-ph].
  %%CITATION = ARXIV:0908.2067;%%







\end{thebibliography}
\end{document}